 \documentclass[12pt]{article}
 \usepackage{amsmath}
 \usepackage{amsfonts}
 \usepackage{amssymb}
 
 \usepackage{pst-plot}
 \usepackage{pstricks}
 \usepackage{multido}
 \usepackage{fp}
 \usepackage{ifthen}
 \usepackage{graphicx}
 \usepackage{url}
 \usepackage{indentfirst}

 \allowdisplaybreaks[4]     

 \renewcommand{\L}{Lema\^{\i}tre}
 \newcommand{\LT}{\L-Tolman}

 \newcommand{\showlabel}[1]{
   \label{#1}
 }
 \newcommand{\pb}[2]{                       
    \parbox[t]{#1}{
       \raggedright
       \setlength{\parskip}{1.2ex}          
       #2
    }
 }
 
 \newcommand{\lra}{\leftrightarrow}
 
 \newcommand{\nn}{\nonumber}
 \newcommand{\er}[1]{(\ref{#1})}          

 \newcounter{paper}
 \newcounter{fg}
 \renewcommand{\thefg}{\arabic{fg}}
 \newcommand{\Fgr}[1]{
   \refstepcounter{fg}
   {\bf Fig.~\thefg}~~
   \label{#1}
 }

 \newcommand{\mb}{\mathbf}
 \newcommand{\bs}[1]{\mbox{\boldmath $#1$}}

 \newcommand{\be}{{\bf e}}     
 \newcommand{\bp}{{\bs{\p}}}     
 \newcommand{\bk}{{\bf k}}     
 \newcommand{\bu}{{\bf u}}     

 \newcommand{\ol}[1]{ \overline{#1} }     

 \newcommand{\Rt}{\dot{R}}

 \newcommand{\vt}{\vartheta}
 \newcommand{\vp}{\varphi}

 \newcommand{\beo}{{\overline{\be}}}
 \newcommand{\eo}{{\overline{e}}}
 \newcommand{\xo}{{\overline{x}}}
 \newcommand{\tao}{{\overline{\tau}}}
 \newcommand{\gmo}{{\overline{\gamma}}}

 \newcommand{\bet}{{\tilde{\be}}}
 \newcommand{\xt}{{\tilde{x}}}
 \newcommand{\tat}{{\tilde{\tau}}}
 \newcommand{\rt}{{\tilde{r}}}
 \newcommand{\vtt}{{\tilde{\vt}}}
 \newcommand{\vpt}{{\tilde{\vp}}}

 \renewcommand{\th}{{\hat{t}}}
 
 \newcommand{\beh}{{\hat{\be}}}
 
 \newcommand{\eh}{{\hat{e}}}
 \newcommand{\xh}{{\hat{x}}}
 \newcommand{\tah}{{\hat{\tau}}}
 \newcommand{\chh}{{\hat{\chi}}}
 \newcommand{\vth}{{\hat{\vt}}}
 \newcommand{\vph}{{\hat{\vp}}}
 
 \newcommand{\nh}{{\hat{n}}}

 \renewcommand{\d}{{\rm d}}     
 \newcommand{\p}{\partial}     
 \newcommand{\td}[2]{\frac{{\rm d} #1}{{\rm d} #2}}     
 \newcommand{\tdil}[2]{{\rm d} #1/{\rm d} #2}     
 \newcommand{\pd}[2]{\frac{\partial #1}{\partial #2} }     
 \newcommand{\ad}[2]{\frac{\delta #1}{\delta #2}}     
 \newcommand{\tdt}[2]{ \frac{{\rm d}^2 #1}{{\rm d} {#2}^2} }     
 \newcommand{\tdtil}[2]{ {\rm d}^2 #1/{\rm d} {#2}^2 }     
 \newcommand{\adt}[2]{ \frac{\delta^2 #1}{\delta {#2}^2} }     

 \title{Calculating Observables in Inhomogeneous Cosmologies I: \\
        General Framework}

 \author{
   Charles Hellaby\thanks{\tt Charles.Hellaby@uct.ac.za}
   \and
   Anthony Walters\thanks{\tt Tony.Walters@uct.ac.za} \\
   {\small \it Dept. of Maths. and Applied Maths,
   University of Cape Town,
   Rondebosch,
   7701,
   South Africa}
 }
 
 \date{}

\begin{document}
 \sffamily


 \maketitle

 \begin{abstract}
   We lay out a general framework for calculating the variation of a set of cosmological observables, down the past null cone of an arbitrarily placed observer, in a given arbitrary inhomogeneous metric.  The observables include redshift, proper motions, area distance and redshift-space density.  Of particular interest are observables that are zero in the spherically symmetric case, such as proper motions.  The algorithm is based on the null geodesic equation and the geodesic deviation equation, and it is tailored to creating a practical numerical implementation.  The algorithm provides a method for tracking which light rays connect moving objects to the observer at successive times.  Our algorithm is applied to the particular case of the Szekeres metric.  A numerical implementation has been created and some results will be presented in a subsequent paper.  Future work will explore the range of possibilities.
 \end{abstract}

 \section{Aim and Motivation}

   On the scale of clusters and superclusters, cosmic flows have been the subject of extensive research, usually based on interpreting observations in the light of Newtonian gravitational dynamics, such as the beautiful graphics of Tully et al \cite{TuCoHoPo14}.  In fact, the ability to measure flows on very large scales may not be far off.  Current radio astrometry, by analysing decades of data, is able to detect proper motions of extra-galactic objects and even quasars, of the order of a few micro-arc-seconds per year \cite{Tito09,Tito10,TiLaGo11,MoFrLaTiBa11}.  Interestingly, Krasinski and Bolejko \cite{KraBol13} did estimate the rate of ``drift across the sky" of sources in inhomogeneous models to be of order $10^{-6}$ arcseconds/year.  

   To date, cosmic flows have not really been investigated as a feature of inhomogeneous cosmological models.  This is due to either the high symmetry of the situations modelled (central observer in a spherically symmetric model) or the complexity of doing the calculation in a non-symmetric model.  We here present a general framework for calculating various observables for abitrarily moving observers and sources in an arbitrary spacetime.  These observables include redshift, proper motion (``cosmic flow"), area distance and redshift space density, amongst other possibilities.  The approach is designed for efficient numerical implementation, and numerical examples will be presented in a subsequent paper, Paper II \cite{WalHel17}.

   An ``inhomogeneous cosmological model" is an exact solution of the Einstein field equations (EFEs) that generically has non-zero density, though regions of vacuum are often possible.  To be realistic, the equation of state should be physically acceptable, even if simplified.  Any given models can be applied on any scales, since the EFEs have a scale freedom.  The study of such models is an important complement to perturbation methods, and Newtonian N-body simulations, since they are fully relativistic and fully non-linear.  As measurements become increasingly accurate, the need for exact models will become more apparent.  Thus they are very useful regardless of whether there is inhomogeneity on the largest observable scales.

   In fact, analyses of observations continue to exhibit ``tensions" which hint at anisotropy and even redshift-dependence in the Hubble flow \cite{%
 LanMag05,%
 InoSil06,%
 WatFelHud09,%
 SyVaBa09a,%
 SyVaBa09b,%
 Tito09,%
 ClCaGrSo12,%
 CHRCSG13,%
 KeeBarCow13,%
 KeeBarCow14,%
 MaAmSaVa13,%
 BatChaMos15,%
 BCKRW16,%
 BeVeRi16,%
 McKWil16,%
 MCCWSG16,%
 PaHyNoHw17,%
 FolKno17,%
 FeeMorDal17,%
 WuHut17,%
 AylEtAl17%
 }.  Whilst these have yet to be solidly confirmed on the large scale, it is certain that the structures that are known to exist must have some effect on the smaller scale.  Similarly the near-alignment of the dipole, quadrupole and octopole moments in the CMB may conceivably be due to the effect of some large scale structure \cite{Vale05}.

   The presence or absence of proper motions on large scales would put constraints on possible models.  The development of non-linear models of the observable effects of different structures would assist in interpreting results and calculating constraints.  Therefore, it is of considerable interest to investigate how inhomogeneities of different shapes and sizes would affect the perceived Hubble flow and the pattern of peculiar velocities.  Perturbation theory is well suited to statistical analysis of fluctuations, whereas the use of exact inhomogeneous models is more appropriate for building specific models and analysing particular observational features.  This is a first exploration of the issue.  

   Our interest here is in determining observations for a given observer in a given spacetime metric.  In other words, we seek to use a known inhomogeneous solution of the EFEs to construct a particular model, then see what the observations would be and how they would change.  One aim is to develop models which have acceptably small fluctuations at recombination, but which have a measurable flow at later times which may or may not have developed into a strong inhomogeneity, and to check the magnitude of the resultant perturbations of the CMB.  Preliminary versions of this work can be found at \cite{HelWalTalks}.

   This is in a sense the reverse of the `Observational Cosmology' (OC) approach \cite{KriSac66,ENMSW85,StElNe92a,StNeEl92b,StNeEl92c,SNME92,AraSto99,AABFS01,ArRoSt01,RibSto03,AlIrRiSt07,ASAB08}, and the related Metric of the Cosmos approach \cite{MuHeEl97,Hel01,Hel03,Hel06,LuHel07,McCHel08,HelAlf09,AlfHel09} or inverse problem \cite{YoKaNa08,Yoo10,TokYoo16}.  There one defines coordinates based on the observer's PNC, with incoming light rays labelled by observed sky angles and time, and the idea is to use observations plus the EFEs to specify what our spacetime metric is.  However these approaches treat our observations as occurring at a single time relative to cosmic timescales, so that the EFEs are needed to evolve the spacetime to the future and past.  If, on the other hand, observations are long enough or accurate enough to detect variation on the large scale, then it is more important that each PNC (originating at successive times on the observer's worldline) be constructed in a similar way.  

   We have recently learned of overlapping work by Korzy\'{n}ski \cite{Korz17}, that also finds formulas for observables using past null geodesics and the geodesic deviation equation, for general sources and observers in general spacetimes.  So far, though, it does not have a numerical implementation, and there are important differences in the approach.

 \section{Mapping the Fluid Flow to Proper Motion Measurements on an Observer's Sky}
 \showlabel{Mapping}

   Determining the evolution of observations in a general metric is a highly non-trivial matter, primarily because light rays from source to observer at successive times will not cross the same set of intervening spatial points.  One can trace a single ray backwards from the observer, and at some point choose the source.  But, for light emitted at a later time from that same source, what is the correct direction to follow to arrive back at the observer, given that the spacetime is non-symmetric and dynamic?  It is essential to know these light paths in order to calculate the evolution of redshift, diameter or luminosity distance, proper motion, etc.  Thus one appears to be saddled with a major numerical exercise.  We here propose a method which will give instantaneous rates of change of observations, such as bulk flow patterns, and which will also serve as a guide to numerical light ray tracking methods for determining the longer term evolution of observations.

   The past null cone (PNC) is the basis of all astronomical measurements.  We consider an observer $O$, with 4-velocity $u^a_o$ who sets up null cone coordinates $\xh^\alpha = (\tau, \chi, \vt, \vp)$ centred on her own position.  We envisage that the spacetime contains inhomogeneities, and that the metric describes these to some level of accuracy.  We do not assume that observer and emitter are each comoving with the local mean flow, though that is the simplest option%
 \footnote{\sf The observer's motion will never be exactly the same as the cosmological flow.  If the metric is a Friedmann-\L-Robertson-Walker (FLRW) approximation, then there are proper motions of $\sim600$~km/s away from the Hubble flow.  If the metric is an exact description down to the level of individual galaxies, say, there would still be the orbit of the sun, the orbit of the earth, and the earth's rotation to account for.  The idea here is to investigate how cosmic flows might be modelled, so we will at present ignore these issues.}%
 .  The set of light rays that arrive at a particular observer $O$ form a 3-parameter congruence of null geodesics with tangent vector $k^a$.  The natural way to label these light rays is with $O$'s right ascension and declination, or with galactic lattitude and longitude, and the time of observation in $O$'s frame.  Anything that $O$ sees in a particular direction will be labelled with the same sky coordinates $(\vt, \vp)$, and the time of observation $\tau$.  

   In order to map emission events in spacetime to measurements of the angle and time as observed, we need to Lie drag the observer's $(\tau, \vt, \vp)$ coordinates down the light rays of her past null cone.  By this construction, $\tau$, $\vt$ \& $\vp$ are constant along each incoming null geodesic.  Although it will be convenient to deal with basis vectors below, note that at this point we are primarily mapping scalars ($\tau$, $\vt$, $\vp$) down the PNCs of observation events, rather than vectors%
 \footnote{\sf Parallel transport is obviously wrong for $\vt$ \& $\vp$}%
 .

   The only natural parameter or ``distance" down the light ray is redshift, $z$, as the true distance of a source and the time light was emitted are not measurable.  However, $z$ is not guaranteed to be monotonically increasing \cite{MBHE98}, and more importantly we will need to integrate down these light rays relative to some parameter.  For this purpose it makes sense to construct the affine parameter $\chi$, thus $k^a = \tdil{x^a}{\chi}$.  To complete the coordinate system, it is sensible to use $\chi$ and $k^b$ as the last coordinate and basis vector.

   We propagate these coordinates as follows, assuming we have a given metric: \\
 ${}$~~~~~~~~~~~~(a) choose a set of reference directions at the observer, and set up \\
 ${}$~~~~~~~~~~~~~~~~ time-radial-angular coordinates $\tat, \rt, \vtt, \vpt$ in the observer's vicinity; \\
 ${}$~~~~~~~~~~~~(b) solve for the PNC --- the paths of the incoming light rays; \\
 ${}$~~~~~~~~~~~~(c) Lie drag the observer's coordinates and basis vectors along the PNC. \\
 We now deal with these in turn.  The remainder of section \ref{Mapping} deals with calculations down a single ray.  Once that is working correctly, extension to many sky directions and many observation times becomes quite easy.  
See Fig 1 for an illustration.  

 {\footnotesize
 \psset{unit=1mm, xunit=1mm, yunit=1mm}
 \pspicture*(0,0)(161,166)     
 \psset{linewidth=0.7pt}
 \psset{linecolor=magenta}
 \pscircle*(11,25){0.7}
 \psellipse[linestyle=dashed](11,25)(150,10) 
 \psset{linewidth=1.2pt}
 \psset{linecolor=red}
 \pscurve(12,0)(11,25)(7,95)(10,135)(11,163)(11,165) 
 \rput[r](9,135){$O$}
 \rput[l](14,5){Observer's worldline}
 \pscurve(96,0)(100,17.1)(120,50) 
 \rput[rt](98.3,16){$E$}
 \rput[lb](120,51){Emitter's worldline}
 \psset{linewidth=0.8pt}
 \psset{linecolor=black}
 \pscurve(10,135)(28,113.42)(46,88.84)(64,63.26)(82,38.68)(100,17.1) 
 \psline[arrowsize=4pt 2,arrowlength=1.6,arrowinset=0.3]{<-}(64,63.26)(65,61.8)
 \pscurve(10,135)(32,113.64)(56,89.28)(76,63.92)(98,39.56)(120,18.2) 
 \psline[arrowsize=4pt 2,arrowlength=1.6,arrowinset=0.3]{<-}(76,63.92)(77,62.8)
 \pscurve(11,163)(28.8,141.4)(46.6,116.8)(64.4,91.2)(82.2,66.6)(100,45) 
 \psline[arrowsize=4pt 2,arrowlength=1.6,arrowinset=0.3]{<-}(64.4,91.2)(65.4,89.8)
 \rput[rt](10,163){$O'$}
 \rput[r](62.7,62){Incoming light rays}
 \psset{linewidth=2pt}
 \psset{linecolor=green}
 \psline{->}(10,135)(11,163)
 \rput[l](12,149){$\mb{u}_o$}
 \psline{->}(100,17.1)(109,37)
 \rput[b](109,38){$\mb{u}_e$}
 \psset{linewidth=2pt}
 \psset{linecolor=blue}
 \psline{->}(28,113.42)(28.8,141.4) 
 \rput[lb](29.8,141.4){$\beh_\tah$}
 \psline{->}(28,113.42)(32,113.64)
 \rput[l](33.5,114){$\beh_\vth$}
 \psline{->}(28,113.42)(36,102.8)
 \rput[rt](36,103){$\mb{k} = \beh_\chh$}
 \rput[rt](27,112.42){$N$}
 \psline{->}(100,17.1)(100,45) 
 \rput[lb](101,45){$\beh_\tah$}
 \psline{->}(100,17.1)(120,18.2)
 \rput[t](118,16){$\beh_\vth$}
 \psline{->}(100,17.1)(110,5.2)
 \rput[lt](110,4){$\mb{k} = \beh_\chh$}
 \rput[lt](83,160){
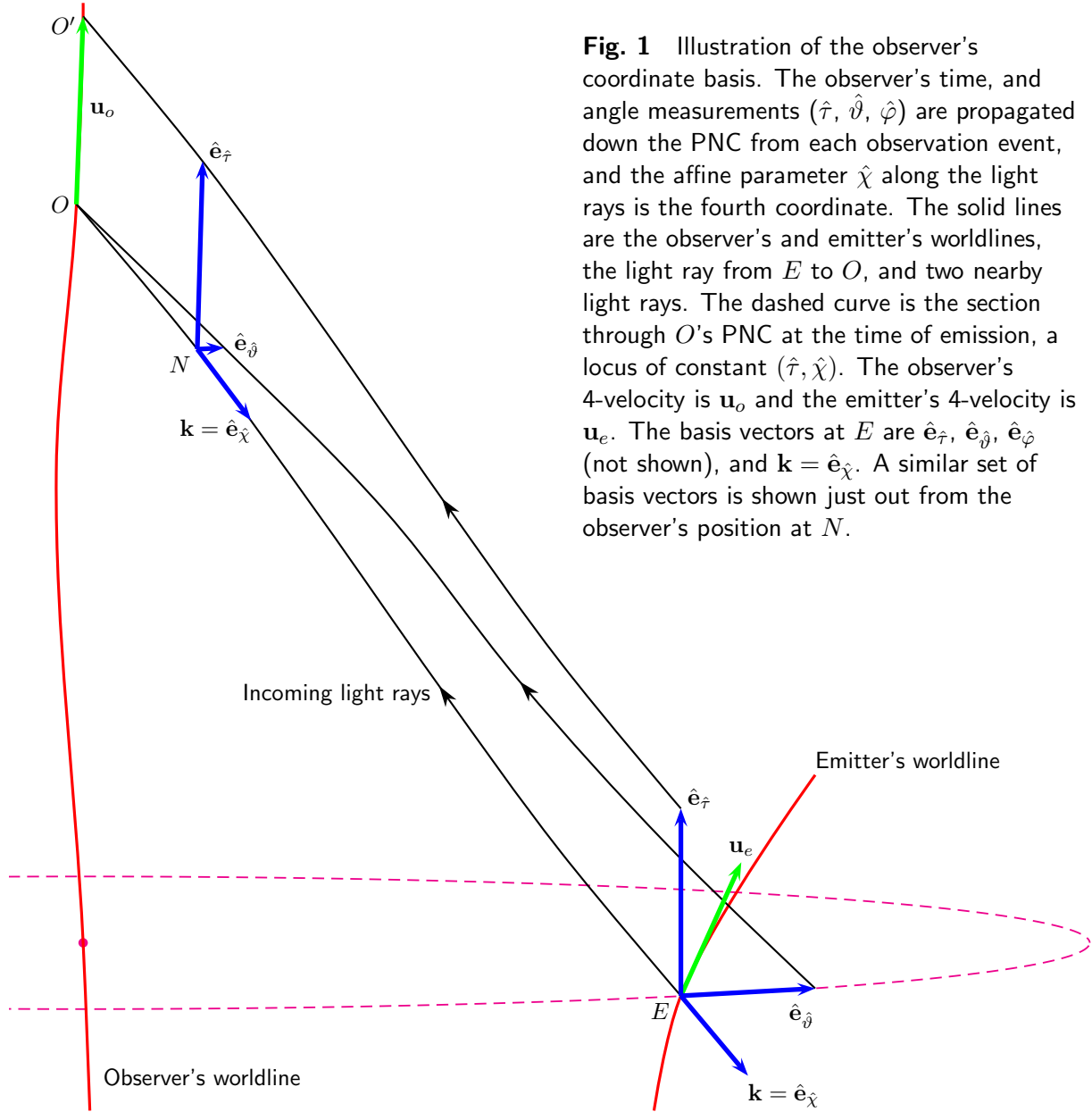
   \pb{74mm}{\sf \small  \Fgr{PNTpic} Illustration of the observer's coordinate basis.  The observer's time, and angle measurements ($\tah$, $\vth$, $\vph$) are propagated down the PNC from each observation event, and the affine parameter $\chh$ along the light rays is the fourth coordinate.  The solid lines are the observer's and emitter's worldlines, the light ray from $E$ to $O$, and two nearby light rays.  The dashed curve is the section through $O$'s PNC at the time of emission, a locus of constant $(\tah, \chh)$.  The observer's 4-velocity is $\mb{u}_o$ and the emitter's 4-velocity is $\mb{u}_e$.  The basis vectors at $E$ are $\beh_\tah$, $\beh_\vth$, $\beh_\vph$ (not shown), and $\mb{k} = \beh_\chh$.  A similar set of basis vectors is shown just out from the observer's position at $N$.}
 }
 \endpspicture
 }

 \subsection{Propagating the Observer's Coordinates and Basis}

 \paragraph{(a) Observer's Sky Coordinates \& Local Basis}
   For the given metric $g_{ab}$ and coordinates $x^c$, we specify the observer's worldline, $O$, and at least locally we set up an orthonormal tetrad (ONT) $\beo_e$ with $\beo_0$ in the direction of the observer's 4-velocity,
 \begin{align}
   \beo_i \big|_o & = \big[ \eo_i{}^a \, \bp_a \big]_0 ~,   \showlabel{OthNmBbar}
 \end{align}
 where $\bp_a$ are the coordinate basis vectors and the components $\ol{e}_i{}^a$ define the new basis $\beo_i$.  We allow the observer's worldline to be a general timelike path; the simplest example is for $O$ to be comoving with the fluid flow (the timelike eigenvector of the matter).  Since it is always possible to set up local coordinates such that the Christoffel symbols are zero to 1st order, there will be associated coordinates $\xo^e$ in the neighbourhood of $O$, and $\xo^0 = \tao$ is the observer's proper time.  Whether or not this is a coordinate basis, we will only need the $\beo_i$ and their first derivatives at $O$, and we will construct a coordinate basis from them.

   Taking the $\beo_1$, $\beo_2$ \& $\beo_3$ basis vectors to be the directions of $(\vt, \vp) = (\pi/2, 0)$, $(\vt, \vp) = (\pi/2, \pi/2)$ \& $\vt = 0$, we convert to a spherical basis%
 \footnote{\sf This basis is of course degenerate at the observer.}%
 , with coordinates $\tilde{x}^i = (\tat, \rt, \vtt, \vpt)$
 \begin{align}
 \begin{split}
   \bet_\tat & = \beo_0 \\
   \bet_\rt & = \sin\vtt \cos\vpt \, \beo_1 + \sin\vtt \sin\vpt \, \beo_2
      + \cos\vtt \, \beo_3 \\
   \bet_\vtt & = \rt \cos\vtt \cos\vpt \, \beo_1 + \rt \cos\vtt \sin\vpt \, \beo_2
      - \rt \sin\vtt \, \beo_3 \\
   \bet_\vpt & = - \rt \sin\vtt \sin\vpt \, \beo_1 + \rt \sin\vtt \cos\vpt \, \beo_2 ~,
 \end{split}
   \showlabel{SphBtilde}
 \end{align}
 where $\rt$ is a proper radius from $O$ and $\bet_\rt$ is a spacelike unit vector pointing outwards from $O$ and orthogonal to $\bet_\tat$, $\bet_\vtt$, $\bet_\vpt$.  Then we convert to a null radial basis vector by writing
 \begin{align}
   \tah = \rt + \tat ~,~~ \chh = \rt ~~~~~~\lra~~~~~~
   \tat = \tah - \chh ~,~~ \rt = \chh ~,
 \end{align}
 obtaining
 a past null tetrad (PNT) for the observer, associated with coordinates $\xh^i = (\tah, \chh, \vth, \vph)$
 \begin{align}
 \begin{split}
   \beh_\tah & = \bet_\tat = \beo_0 \\
   \beh_\chh & = - \bet_\tat + \bet_\rt = - \beo_0 + \sin\vth \cos\vph \, \beo_1
      + \sin\vth \sin\vph \, \beo_2 + \cos\vth \, \beo_3 \\
   \beh_\vth & = \bet_\vtt = \chh \cos\vth \cos\vph \, \beo_1
      + \chh \cos\vth \sin\vph \, \beo_2 - \chh \sin\vth \, \beo_3 \\
   \beh_\vph & = \bet_\vpt = - \chh \sin\vth \sin\vph \, \beo_1
      + \chh \sin\vth \cos\vph \, \beo_2 ~,
 \end{split}
   \showlabel{NlRdBhat}
 \end{align}
 and these are the basis vectors that are to be propagated down the observer's PNC.  Note that $\tah$ is a null coordinate --- it labels particular past null cones%
 \footnote{\sf This is different from the ``fluid-ray tetrad" of \cite{SNME92}, since the latter uses the local fluid 4-velocity.}%
 .

   For clarity, we summarise our conventions for these frames below.  Note that the ONT ``coordinates" $\xo^i$ and the observer's natural ``coordinates" $\xt^m$ will only exist locally.
 \[
 \begin{tabular}{|l|l|l|l|}
   \hline
   \rule[-0.3ex]{0em}{2.7ex} Coord System/Basis & ``Coordinates" & Basis & Index Symbols \\ \hline \hline
   \rule[-0.3ex]{0em}{2.7ex} coordinates of given metric & $x^a$ & $\bp_b$ & $a, b, c, \cdots$ \\ \hline
   \rule[-0.3ex]{0em}{2.7ex} ONT near $O$ & $\xo^i$ & $\beo_j$ & $i, j, k, \cdots$ \\ \hline
   \rule[-0.3ex]{0em}{2.7ex} observer's natural coords near $O$ & $\xt^m = (\tat, \rt, \vtt, \vpt)$ & $\bet_n$ & $m, n, o, \cdots$ \\ \hline
   \rule[-0.3ex]{0em}{2.7ex} observer's PNC coords \& PNT  & $\xh^\alpha = (\tah, \chh, \vth, \vph)$ & $\beh_\beta$ & $\alpha, \beta, \gamma, \cdots$ \\ \hline
 \end{tabular}
 \]

 \paragraph{(b) Propagating $k^a$}
   The observer's PNC is propagated outwards by means of the null geodesic equation
 \begin{align}
   \ad{k^a}{\chh} & = 0   \showlabel{kNullGeodAbs}
   \intertext{or rather}
   \td{k^a}{\chh} & = - k^b \Gamma^a{}_{bc} k^c ~,~~~~ k^a k_a = 0 ~,~~~~ \td{x^a}{\chi} = k^a   \showlabel{kNullGeod}
 \end{align}
 which is solved numerically, giving the path of the light ray, the affine parameter $\chh$, and the basis vector $k^b = \eh_\chh{}^b$.  Obviously, we need a starting point $[x^a]_o$ and an initial $[k^a]_o$.  The latter follows from \er{NlRdBhat} and the choice of ray direction, $(\vth, \vph)$, via
 \begin{align}
   [\bk]_o & = \beh_\chh
   = - \beo_0 + \sin\vth \cos\vph \, \beo_1 + \sin\vth \sin\vph \, \beo_2 + \cos\vth \, \beo_3 ~.   \showlabel{inik}
 \end{align}
 This ensures that, in $O$'s orthonormal frame, the initial null vector is $k^\alpha = (-1, 1, 0, 0)$, or in terms of coordinate components,
 \begin{align}
   |k^b u_b u^a|_o = 1 = |k^a (\delta^c_a + u^c u_a)|_o ~~~~\to~~~~ (k^b u_b)_o^2 ~.   \showlabel{inikc}
 \end{align}
 This can be made more explicit once a metric is chosen --- see later in \S\ref{SzkMdl}.

 \paragraph{(c) Propagating the Past Null Tetrad}
   Beyond the immediate neighbourhood of $O$, we define the observer's PNC coordinates and basis by Lie dragging them down the PNC.  Thus in principle, the basis could be propagated by,
 \begin{align}
   [\be_\chh, \be_\alpha] = {\cal L}_{\bf k} \beh_\alpha & = 0
   = k^b \p_b \eh_\alpha{}^a - \eh_\alpha{}^b \p_b k^a   \showlabel{ekLie} \\
   \to~~~~~~~~ \td{}{\chh} \eh_\alpha{}^a & = \eh_\alpha{}^b \p_b k^a ~.   \showlabel{dedchiLie}
 \end{align}
 However, this requires knowledge of the transverse derivative, $\p_b k^a$, which would be an extra numerical calculation.  Alternatively, since the $\eh_\beta{}^c$ obey \er{ekLie} and $k^d$ obeys \er{kNullGeod}, the geodesic deviation equation holds,
 \begin{align}
   \adt{\eh_\alpha{}^a}{\chh} & = - R^a{}_{bcd} \, k^b \, \eh_\alpha{}^c \, k^d ~,   \showlabel{GeodDev}
 \end{align}
 and so the basis vectors $\{ \beh_\tah$, $\beh_\vth$, $\beh_\vph \}$ are deviation vectors linking nearby light rays.  Typically, the Christoffel symbols $\Gamma^a{}_{bc}$ and Riemann tensor components $R^a{}_{bcd}$ are available analytically for a given metric.  However, we avoid the transverse derivative in \er{dedchiLie} at the cost of having to solve a higher order differential equation (DE).  Just as \er{kNullGeod} is preferable to \er{kNullGeodAbs} numerically, so numerical integration of the components $\eh_\alpha{}^a$ along $k^b$ requires that \er{GeodDev} be converted to an equation for $\tdtil{\eh_\alpha{}^c}{\chh}$ --- see appendix \ref{GDprop}.   Therefore putting $V^a = k^a$, $v = \chh$ and $W^b = \eh_\alpha{}^b$ into \er{WvvUp}, the equation to be integrated is
 \begin{align}
   \tdt{\eh_\alpha{}^a}{\chh}& = - k^b \left( 2 \Gamma^a{}_{bc} \td{\eh_\alpha{}^c}{\chh}
      + \eh_\alpha{}^c k^d \Gamma^a{}_{db,c} \right) ~.   \showlabel{GeodDevTot}
 \end{align}
 For this purpose we need initial conditions on $\eh_\alpha{}^a$ and $\tdil{\eh_\alpha{}^a}{\chh}$ at $O$, which we derive in the next section.
We establish under what conditions the resulting basis is a coordinate basis in appendix \ref{Cmt}.

   Once the observer's coordinates have been propagated down the PNC, they provide the transformation between the metric coordinates and the observer's PNC coordinates,
 \begin{align}
   \eh^\alpha{}_c = e_c{}^\alpha = \pd{\xh^\alpha}{x^c} ~,~~~~~~
   e^c{}_\alpha{} = \eh_\alpha{}^c = \pd{x^c}{\xh^\alpha} ~.
 \end{align}
 We will see below that what we actually need not just $\eh_\alpha{}^a$ but also its inverse $\eh^\alpha{}_a$, obtained via
 \begin{align}
   \eh^\alpha{}_a \, \eh_\beta{}^a = \delta^\alpha_\beta ~~~~\lra~~~~
   \eh^\alpha{}_b \, \eh_\alpha{}^a = \delta^a_b ~.   \showlabel{invbv}
 \end{align}
 One may therefore consider propagating the dual basis vectors instead, and the DE for this is given in \er{WvvDn}.

   In principle we now have 4 numerically distinct methods of propagating the observer's basis, which should all agree: (i) Lie dragging then inversion, (ii) geodesic deviation then inversion, (iii) inversion then Lie dragging, (iv) inversion then geodesic deviation.  The method pursued below and in Paper II is (ii).

 \subsection{Initial Conditions for the Basis Propagation}
 \showlabel{ICs}

   To solve \er{GeodDevTot} requires initial conditions on the observer's PNC basis and their derivatives, at $O$, i.e. when $\chi \to 0$.  For $\be_\tah$, we require that it coincide with $\bu_o$ at $O$.  
 \begin{align}
   \lim_{\chh \to 0} e_\tah{}^c = u_o^c ~.   \showlabel{etahIC}
 \end{align}
 The observer-origin limits of the third and fourth of \er{NlRdBhat} give us
 \begin{align}
   \beh_\vth \big|_{\chh \to 0} & = \lim_{\chh \to 0} \left[
         \chh \cos\vth \cos\vph \, \beo_1 + \chh \cos\vth \sin\vph \, \beo_2 - \chh \sin\vth \, \beo_3 \right] \nn \\
   \beh_\vth \big|_{\chh = 0} & = 0   \showlabel{evthIC} \\
   \beh_\vph \big|_{\chh \to 0} & = \lim_{\chh \to 0} \left[
         - \chh \sin\vth \sin\vph \, \beo_1 + \chh \sin\vth \cos\vph \, \beo_2 \right] \nn \\
   \beh_\vph \big|_{\chh = 0} & = 0   \showlabel{evphIC}
 \end{align}

   Next consider the $\bk$ vectors at $O$ and $O'$ in fig \ref{PNTpic}.  Once $\bk$ at $O$ is chosen, the one at $O'$ must be in the same direction, as perceived by the observer.  Therefore it is constructed by Fermi transport of $\bk$ along $\bu_o$
 \begin{align}
   \ad{k^a}{\tau} \Bigg|_{\chh = 0} & = \Big[ u_o^b \nabla_b k^a
   - k_b a_o^b u_o^a + k_b u_o^b a_o^a \Big]_{\chh = 0} = 0 ~,
 \end{align}
 where the observer's proper acceleration is
 \begin{align}
   a_o^c & = u_o^d \nabla_d u_o^c = \Big[ \eh_\tah{}^d \nabla_d \eh_\tah{}^c \Big]_{\chh = 0} ~.
 \end{align}
 Thus, using \er{etahIC} and \er{dedchiLie}, we find
 \begin{align}
 \eh_\tah{}^b \p_b k^a + \eh_\tah{}^b \Gamma^a{}_{bc} k^c & = k_b a_o^b \eh_\tah{}^a - k_b \eh_\tah{}^b a_o^a \\
   \to~~~~~~~~ \td{\eh_\tah{}^a}{\chi} \Bigg|_{\chh = 0} & = \Big[ - \eh_\tah{}^b \Gamma^a{}_{bc} k^c
   + k_b a_o^b \eh_\tah{}^a - k_b \eh_\tah{}^b a_o^a \Big]_{\chh = 0} ~.
      \showlabel{detahIC}
 \end{align}
 If the observer is geodesic, then $a_o^b = 0$ and Fermi transport becomes parallel transport.
 Looking at the second of \er{NlRdBhat}, and noting that
 \begin{align}
   \lim_{\chh \to 0} \td{\beo_j}{\vth} & = 0 = \lim_{\chh \to 0} \td{\beo_j}{\vph} ~,
 \end{align}
 since the variation happens at one point, we obtain
 \begin{align}
   \td{\bk}{\vth} \Bigg|_{\chh \to 0} & = \lim_{\chh \to 0} \left[
      \cos\vth \cos\vph \, \beo_1 + \cos\vth \sin\vph \, \beo_2 - \sin\vth \, \beo_3 \right] \\
   \td{\bk}{\vph} \Bigg|_{\chh \to 0} & = \lim_{\chh \to 0} \left[
      - \sin\vth \sin\vph \, \beo_1 + \sin\vth \cos\vph \, \beo_2 \right] ~.
 \end{align}
 Again by \er{dedchiLie} we arrive at
 \begin{align}
   \td{e_\vth{}^a}{\chh} \Bigg|_{\chh \to 0} & = \td{k^a}{\vth} \Bigg|_{\chh \to 0} = \left[
      \cos\vth \cos\vph \, \eo_1{}^a + \cos\vth \sin\vph \, \eo_2{}^a - \sin\vth \, \eo_3{}^a \right]_{\chh = 0}
      \showlabel{devthIC}
 \intertext{and}
   \td{e_\vph{}^a}{\chh} \Bigg|_{\chh \to 0} & = \td{k^a}{\vph} \Bigg|_{\chh \to 0} = \left[
      - \sin\vth \sin\vph \, \eo_1{}^a + \sin\vth \cos\vph \, \eo_2{}^a \right]_{\chh = 0} ~.
      \showlabel{devphIC}
 \end{align}

   Our set of initial conditions are \er{etahIC}, \er{evthIC}, \er{evphIC}, \er{detahIC}, \er{devthIC} and \er{devphIC}.  These will be applied to a specific example metric and PNT below in section \ref{SzkMdl}.

   For the inverse basis propagation DEs \er{WvvDn}, the initial conditions are given by \er{invbv} at $\chi \to 0$ and
 \begin{align}
    \td{}{\chi} \eh^\beta{}_b \Bigg|_{\chh \to 0} =
       \left[ - \eh^\alpha{}_b \, \eh^\beta{}_a \td{}{\chi} \eh_\alpha{}^a \right]_{\chh \to 0} ~,~~~~~~
       \beta = \tah, \vth, \vph ~.
 \end{align}

 \subsection{Proper Motions and Redshift}

   The proper motion of a source, is the rate of change of observed angle with respect to observer time.  We write
 \begin{align}
   \left. \td{\tilde{x}^m}{\tat} \right|_o & = \left[ \pd{\tilde{x}^m}{\xh^\beta}
      \td{\xh^\beta}{\tah} \td{\tah}{\tat} \right]_o
      = \left[ \eh_\beta{}^m \td{\xh^\beta}{\tah} \right]_o ~, \showlabel{dxdtautilde} \\
   \mbox{where}~~~~ \left[ \td{\xh^\beta}{\tah} \right]_o & = \left[ \td{\xh^\beta}{\tah} \right]_e
      = \left[ \pd{\xh^\beta}{x^a} \td{x^a}{\tau_e} \td{\tau_e}{\tah} \right]_e
      = \frac{\big[ \eh^\beta{}_a u^a \big]_e}{(1 + z)} ~,   \showlabel{dxdtauhat}
 \end{align}
 where $\tau_e$ is the source proper time, $\tilde{\tau}_o$ is the observer's proper time, and $\tah$ is its extension down the PNC.  Clearly, the basis vectors, once propagated, have to be inverted numerically, using \er{invbv}.

   In particular, referring to \er{NlRdBhat}, these two equations provide us with the redshift and the components of the proper motion.  From \er{dxdtauhat},
 \begin{align}
   1 = \left. \td{\tah}{\tah} \right|_o & = \frac{\big[ \eh^\tah{}_a u^a \big]_e}{(1 + z)} \nn \\
   \to~~~~~~~~ 1 + z & = \big[ \eh^\tah{}_a u^a \big]_e ~,   \showlabel{redshift}
 \intertext{so from \er{dxdtautilde},}
   \left. \td{\vtt}{\tat} \right|_o & = \frac{\big[ \eh^\vth{}_a u^a \big]_e}{(1 + z)}   \showlabel{appmotth} \\
   \left. \td{\vpt}{\tat} \right|_o & = \frac{\big[ \eh^\vph{}_a u^a \big]_e}{(1 + z)} ~.   \showlabel{appmotph}
 \end{align}
 Naturally, $\tdil{\rt}{\tat} \big|_o = 0$ by construction.

 \subsection{Objective Plane}

   Given a light ray tangent vector $k^a$, and a unit vector defining a chosen time direction $t^c$, then the spatial direction of the light ray, in the spatial frame of $t^c$ is
 \\[2mm]
 {\footnotesize
 \psset{unit=1mm, xunit=1mm, yunit=1mm}
 \pspicture*(-36,-1)(135,71)     
 \psset{linewidth=0.7pt,linecolor=black,linearc=0.1}
 \pspolygon(0,0)(16,10)(16,70)(0,60) 
 \rput[lt](17.3,70){3-surface orthogonal to $v^a$}
 \pspolygon(-30,34)(-14,44)(46,36)(30,26)
 \rput[lt](30,25.5){3-surface orthogonal to $t^a$}
 \psline[linestyle=dashed](0,30)(16,40) 
 \rput[lb](17,41){2-surface orthogonal to both $v^a$ \& $t^b$}
 \psline[linestyle=dotted]{<-}(-22,69)(38,1) 
 \rput[lt](-19,69){ray}
 \psset{linewidth=1.2pt,linecolor=blue}
 \psline{->}(8,35)(8,50)
 \rput[rt](6.7,50){$\th^a$}
 \psline{->}(8,35)(23,33)
 \rput[rb](23,34.5){$v^a$}
 \psline{->}(8,35)(23,18)
 \rput[rt](22.7,17.7){$k^a$}
 \psline[linestyle=dashed,linewidth=0.7pt](23,33)(23,18)
 \rput[lt](85,60){
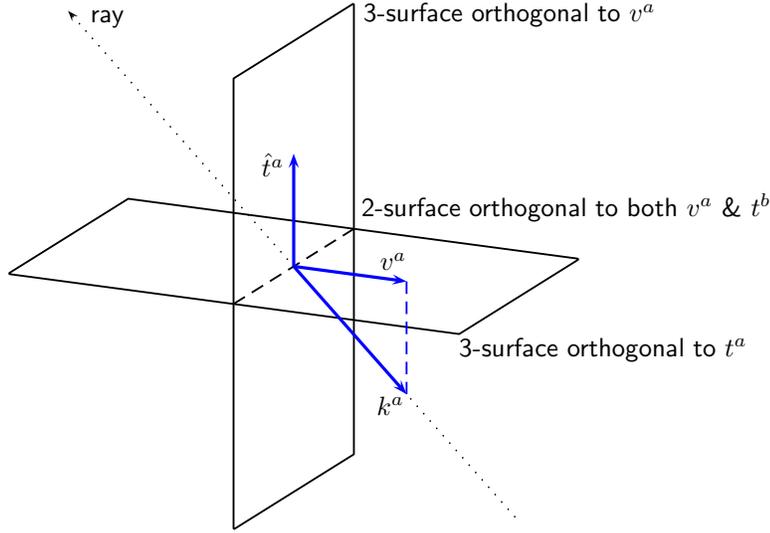
   \pb{45mm}{\sf \small \Fgr{ObjPlane} Illustration of the projection orthogonal to the ray direction and a preferred time direction.  The dashed line represents the image plane with one dimension suppressed.}
 }
 \endpspicture
 }
 \begin{align}
   v^a & = \frac{k^a}{C} + t^a ~,~~~~~~ \mbox{where}~~ C = k_c t^c ~,~~~~~~ C < 0 ~,
 \end{align}
 since it obeys
 \begin{align}
   v^a v_a & = +1 ~,~~~~ v_b t^b = 0 ~~~~\mbox{and hence}~~~~ v^c k_c = C ~.
 \end{align}
 An illustration of this is shown in fig \ref{ObjPlane}.  The tensors
 \begin{align}
   h^a_b & = \delta^a_b + t^a t_b ~,~~~~~~ j^b_c = \delta^b_c - v^b v_c
      \showlabel{h-j-defs}
 \end{align}
 project orthogonally to $t^c$ and $v^d$ respectively, and projecting orthogonally to both gives
 \begin{align}
   h^a_b j^b_c & = \delta^a_c - \frac{k^a k_c}{C^2} - \frac{(t^a k_c + k^a t_c)}{C} ~.
      \showlabel{hj-proj}
 \end{align}
 This defines the emitter's objective plane, but we must select the most appropriate $t^a$.  Now $\beh_\tah$ has been propagated by Lie dragging $\bu_o$ along $\bk$, and so is not necessarily expected to preserve direction in any sense.  Nevertheless, $\beh_\tah$ does represent the separation of two emission events that are perceived by the observer to be from the same direction at subsequent times.  From this point of view, 
 \begin{align}
   t^b = \left[ \frac{\eh_\tah{}^b}{\sqrt{- \eh_\tah{}^c g_{cd} \eh_\tah{}^d}\;} \right]_e ~.
 \end{align}
 It might alternatively be suggested that $t^a$ ought to be the parallel propagation of $\bu_o$ along $\bk$, viz
 \begin{align}
   k^b \nabla_b t^a = 0 ~,~~~~~~ t^c \big|_o = u^c_o ~,
 \end{align}
 though a clear physical argument for this escapes us.
 However, if we say that the diameter or area distance is the ratio of the physical size of the object when you're right next to it, over the measured angular size, then we should choose
 \begin{align}
   t^b = u_e^b ~,   \showlabel{ChoiceOfta}
 \end{align}
 for an object that is comoving with the fluid flow. For cosmologists, this last is how it is normally represented, though only its practical application in FLRW or spherically symmetric (i.e. LT) spacetimes has been looked at.  

 \subsection{Basis Directions Compared with line of sight}
 \showlabel{BasisDirnLnSt}

   In an inhomogeneous model, the propagated angular basis vectors, $\beh_\vth{}^a$ \& $\beh_\vph{}^b$, may not remain perpendicular to the line of sight, and they would develop non-zero values for the dot products
 \begin{align}
   \eh_\beta{}^a \, v_a ~~~~~~\mbox{and}~~~~~~ \eh_\beta{}^c \, \th_c ~,~~~~~~ \beta = \vth, \vph ~,
   \showlabel{edotvt}
 \end{align}
 which have to be compared with their magnitudes,
 \begin{align}
   \sin\psi_\beta = \frac{\eh_\beta{}^a \, v_a}{\sqrt{\eh_\beta{}_b \, \eh_\beta{}^b}\;} ~~~~~~\mbox{and}~~~~~~
   \sinh\xi_\beta = \frac{\eh_\beta{}^c \, \th_c}{\sqrt{\eh_\beta{}_b \, \eh_\beta{}^b}\;} ~,~~~~~~ \beta = \vth, \vph ~.
   \showlabel{edotvt-sin}
 \end{align}
 Calculation of these quantities would indicate how much the angular basis vectors had become ``disoriented".

 \subsection{Area Distance}

   Consider a narrow bundle of incoming light rays emitted by some finite sized object, and subtending observer angles $\d\vth$ and $\d\vph$.  The solid angle of this bundle (at the observer) is
 \begin{align}
   \d\Omega_o = \sin\vth \, \d\vth \, \d\vph ~.
 \end{align}
 As usual, we assume the width of the bundle is everywhere small compared with the cosmological or curvature scales.  Out at the emitter, an affine distance $\chi_e$ or redshift $z_e$ away, $\d\vth$ maps to a physical displacement vector $\mb{dx}^\vth = \beh_\vth \, \d\vth$, and $\d\vph$ maps to $\mb{dx}^\vph = \beh_\vph \, \d\vph$.  These are the relative displacements between the points reached by the rays at $\chi = \chi_e$.  A first estimate of the physical area spanned by these vectors would be obtained via the magnitude of the wedge product $\mb{dx}^\vth \wedge \mb{dx}^\vph$,
 \begin{align}
   \d A = \sqrt{ (g_{ab} \, \eh_\vth{}^a \, \eh_\vth{}^b \, \d\vth^2)(g_{cd} \, \eh_\vph{}^c \, \eh_\vph{}^d \, \d\vph^2)
      - (g_{ab} \, \eh_\vth{}^a \, \eh_\vph{}^b \, \d\vth \, \d\vth)^2 }\; ~.
      \showlabel{dAreaWedge}
 \end{align}
 However, as noted in section \ref{BasisDirnLnSt} and illustrated in fig \ref{ObjRays}, the vectors $\mb{dx}^\vth$ \& $\mb{dx}^\vph$ do not necessarily lie in the objective 2-space that's orthogonal to the light rays and the local matter flow.  
 \\[2mm]
 {\footnotesize
 \psset{unit=1mm, xunit=1mm, yunit=1mm}
 \pspicture*(-8,4)(159,69)     
 \psset{linewidth=0.7pt}
 \psset{linecolor=black}
 \psline{<-}(10,7.2)(105,7.2)
 \psline{<-}(10,59.4)(95,59.4)
 \psset{linecolor=red}
 \rput{40}(60,33.3){
   \psellipse[fillstyle=solid,fillcolor=pink](0,0)(39,10)
   \rput[textcolor=red](0,0){Emitter}
 }
 \pscircle*(86.6,59.4){0.5}
 \pscircle*(33.2,7.2){0.5}
 \psset{linecolor=blue}
 \psline[linewidth=1.5pt]{|-|}(41,7)(41,59.6)
 \rput[r](40,45){$\d A$}
 \psset{linecolor=green}
 \psline[linewidth=1.5pt]{|-|}(53.3,7.0)(45.4,59.6)
 \psset{linestyle=dashed}
 \psline(15,62)(25,-4)
 \psline(30,62)(40,-4)
 \psline(45,62)(55,-4)
 \psline(60,62)(70,-4)
 \psline(75,62)(85,-4)
 \psline(90,62)(100,-4)
 \rput[b](55,64){Lines of constant $\chh$}
 \rput[b]{90}(8,33.3){Ray directions, constant $\vth$, $\vph$, $\tah$}
 \rput[lt](110,66){
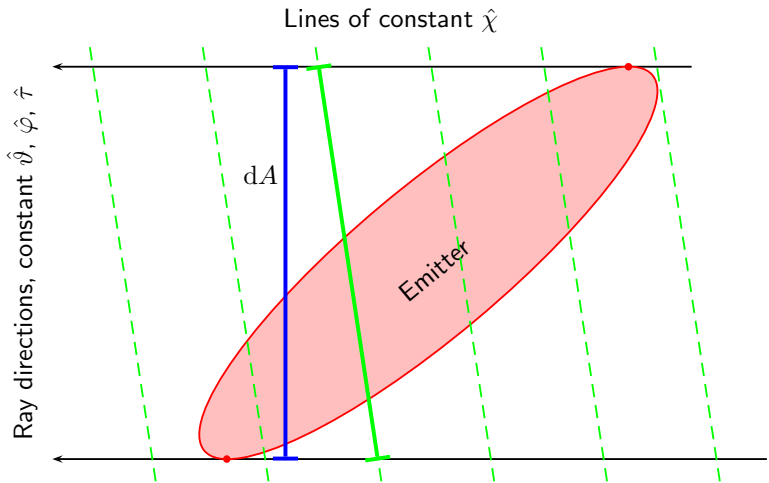
   \pb{45mm}{\sf \small \Fgr{ObjRays} Illustration of the physical extent of the emitter, projected onto the $(\vth,\vph)$ plane (green bar), which lies in the constant $\chh$ surfaces.  Since the constant $\chh$ curves (dashed green lines) are not necessarily orthogonal to the light rays, the correct physical area to use is the projection orthogonal to the line of sight (blue bar).}
 }
 \endpspicture
 }
 \\
 To get the correct area distance, we need the area as projected perpendicular to the line of sight.  We could project each of the displacements before taking their wedge product, 
 \begin{align}
   D^a = \big[ h^a_b j^b_c \d x^c \big]_e ~.   \showlabel{PrjPhysDispl}
 \end{align}
 However, we rather use the permutation pseudo tensor to find the area enclosed by the ray bundle in the objective 2-space,
 \begin{align}
   \d A & = \Big| \eta_{abcd} \, t^a \, v^b \, \big( \eh_\vth{}^c \, \d\vth \big) \big( \eh_\vph{}^d \, \d\vph \big) \Big| \nn \\
   & = \Big| \sqrt{|g|}\; \, \varepsilon_{abcd} \, t^a \left( \frac{k^b}{C} + t^b \right)
      \eh_\vth{}^c \, \eh_\vph{}^d \, \d\vth \, \d\vph \Big| ~,~~~~~~ C = k^c \, t_c \nn \\
   & = \Big| \sqrt{|g|}\; \, \varepsilon_{abcd} \, t^a \left( \frac{k^b}{C} \right)
      \eh_\vth{}^c \, \eh_\vph{}^d \, \d\vth \, \d\vph \Big| ~.
      \showlabel{dArea}
 \end{align}
 The area distance is then
 \begin{align}
   d_A^2 & = \frac{\d A}{\d\Omega_o} = \left| \frac{\sqrt{|g|}\;}{C \sin\vth} \, \varepsilon_{abcd}
      \, t^a \, k^b \, \eh_\vth{}^c \, \eh_\vph{}^d \right| ~,
      \showlabel{dA2}
 \end{align}
 and $t^b = u^b_e$ is the preferred choice.

 \subsection{Number Counts}

   We wish to relate the density of the given spacetime metric to the observed number counts in redshift space.  Consider an element of the PNC 3-surfaces of dimensions $\d\vth$, $\d\vph$, $\d\chh$, as depicted in fig \ref{NmCnts}, and suppose that $N$ sources are observed within that `volume', having average mass $\mu$.  In principle, $\mu$ could be a function of position and time.

 {\footnotesize
 \psset{unit=0.8mm, xunit=0.8mm, yunit=0.8mm}
 \pspicture*(4,11)(146,148)     
 \psset{linewidth=0.7pt}
 \psset{linecolor=cyan}
 \psellipticarc[linestyle=solid](11,45)(120,8){354}{357}                            
 \rput[r](80,39){\pb{10mm}{$\chh$\\$z$}}
 \psellipticarc[linestyle=solid](11,25)(150,10){354}{357}                     
 \rput[lb](122,20){\pb{10mm}{$\chh + \d\chh$\\$z + \d z$}}
 \psset{linewidth=0.8pt}                                                         
 \psset{linecolor=black}
 \pscurve(10,135)(28,113.42)(46,88.84)(64,63.26)(82,38.68)(100,17.1)
 \psline[arrowsize=4pt 2,arrowlength=1.6,arrowinset=0.3]{<-}(64,63.26)(65,61.8)
 \psline[linewidth=0.5pt,linestyle=dashed,dash=0.5 1](82.04,38.65)(82.04,20.7)      
 \psline[linewidth=0.5pt,linestyle=dashed,dash=0.5 1](97.98,39.6)(97.98,21.5)
 \pspolygon[linecolor=blue,linestyle=dashed,dash=0.8 0.6,linearc=0.02](82.04,20.7)(100,17.1)(119.9,18.2)(97.98,21.5)
 \rput[r](80,20){$\d^3V$}
 \rput[r](62.7,62){Incoming light rays}
 \psset{linewidth=0.7pt}                            
 \psset{linecolor=red}
 \rput[r](9,135){$O$}
 \psline[dotsize=0 2]{*-}(87,20.5)(87,60.5)
 \psline[dotsize=0 2]{*-}(90,19.7)(90,59.7)
 \psline[dotsize=0 2]{*-}(92,20.4)(92,60.4)
 \psline[dotsize=0 2]{*-}(94,19.2)(94,59.2)
 \psline[dotsize=0 2]{*-}(97,19.3)(97,59.3)
 \psline[dotsize=0 2]{*-}(99,20.7)(99,60.7)
 \psline[dotsize=0 2]{*-}(100,17.1)(100,57.1)
 \psline[dotsize=0 2]{*-}(105,19.5)(105,59.5)
 \psline[dotsize=0 2]{*-}(103,18.3)(103,58.3)
 \psline[dotsize=0 2]{*-}(107,18)(107,58)
 \psline[dotsize=0 2]{*-}(111,19)(111,59)
 \psline[dotsize=0 2]{*-}(115,18.5)(115,58.5)
 \rput[t](100,16){$E$}
 \rput[b](100,63){Source worldlines}
 \psset{linewidth=2pt}
 \psset{linecolor=green}
 \psline{->}(10,135)(6,145)
 \rput[l](8,145){$\mb{u}_o$}
 \psline{->}(100,17.1)(100,27)
 \rput[lt](101.3,27){$\mb{u}_e$}
 \psset{linewidth=0.8pt}                                
 \psset{linecolor=black}
 \pscurve(10,135)(32,113.64)(56,89.28)(76,63.92)(98,39.56)(120,18.2) 
 \psline[arrowsize=4pt 2,arrowlength=1.6,arrowinset=0.3]{<-}(76,63.92)(77,62.8)
 \rput[lt](83,140){
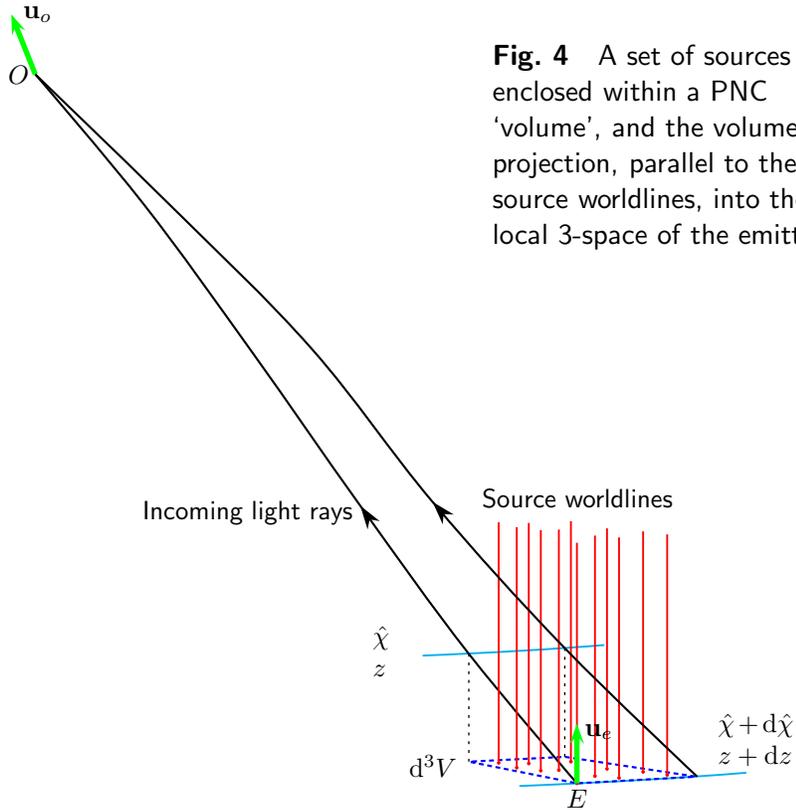
   \pb{50mm}{\sf \small  \Fgr{NmCnts} A set of sources enclosed within a PNC `volume', and the volume's projection, parallel to the source worldlines, into the local 3-space of the emitters.}
 }
 \endpspicture
 }
 \\
 The redshift-space number density, in units of number per unit redshift per unit solid angle, is
 \begin{align}
   \nh = \frac{N}{\d z \, \sin\vth \, \d\vth \, \d\vph}
 \end{align}
 so the mass within the region is
 \begin{align}
   \d M = \mu \, \nh \, \sin\vth \, \d\vth \, \d\vph \, \d z ~.
 \end{align}
 In this subsection, all quantities are evaluated at the emitter.

   The physical 3-volume containing those sources is the projection of this null slice perpendicular to the local emitter 4-velocity $u^a_e$,
 \begin{align}
   \d^3V & = \left| \eta_{abcd} \, u^a_e \, (\eh_\chh{}^b \, \d \chi) \, (\eh_\vth{}^c \, \d \vth)
      \, (\eh_\vph{}^d \, \d \vph) \right| \\
   & = \left| \sqrt{|g|}\; \epsilon_{abcd} \, u^a_e \, k^b \, \eh_\vth{}^c \, \eh_\vph{}^d \, \d \chi \, \d \vth
      \, \d \vph \right| \\
   & = \frac{\d A \, \d\chh}{|C|} ~,
 \end{align}
 using $\d A$ from \er{dArea}.  If the metric density is $\rho$, then the mass in this volume element is
 \begin{align}
   \d M & = \rho \, \d^3V ~.
 \end{align}

   By \er{redshift}, the redshift displacement corresponding to $\d\chh$ is
 \begin{align}
   \d z = \td{z}{\chh} \d\chh & = \left( u^a_e \, \td{}{\chh} \eh^\tah{}_a + \eh^\tah{}_a \td{}{\chh} u^a_e \right) \d\chh ~,
   \showlabel{dzdchiGen}
 \end{align}
 which for comoving emitters is
 \begin{align}
   \d z = u^a_e \left( \td{}{\chh} \eh^\tah{}_a \right) \d\chh ~.
   \showlabel{dzdchiGenCMv}
 \end{align}
 For non-comoving emitters, one would need to define their 4-velocity as a vector field, in order to calculate \er{dzdchiGen}.  Putting these expressions together, we find
 \begin{align}
   \d M & = \rho \left| \sqrt{|g|}\; \epsilon_{abcd} \, u^a_e \, k^b \, \eh_\vth{}^b \, \eh_\vph{}^b \right|
      \d \chi \, \d \vth \, \d \vph \\
   & = \mu \, \nh \, \sin\vth \, \d\vth \, \d\vph \, \d z \\
   \to~~~~~~~~ \nh & = \frac{\rho \left| \sqrt{|g|}\; \epsilon_{abcd} \, u^a_e \, k^b \, \eh_\vth{}^b \, \eh_\vph{}^b \right|}
                            {\mu \, \sin\vth \left( \td{z}{\chh} \right)} ~.
      \showlabel{nhat}
 \end{align}
 Eq \er{nhat} allows numerical computation of the observable number density in redshift space, given the metric quantities and $\mu$.

 \section{The Szekeres Model}
 \showlabel{SzkMdl}

 \subsection{Basics}
 \showlabel{SzkBas}

   We now apply the above general framework to the case of the Szekeres (S) metric \cite{Szek75a,Szek75b}.  This metric is interesting because it is an inhomogneous dust cosmology with 5 physical arbitrary functions of one variable and no symmetries (Killing vectors).  There are two families --- Datt-Kantowski-Sachs-like (``$\beta' = 0$") and \LT-Ellis-like (``$\beta' \neq 0$"), but the former can be written as a regular limit of the latter \cite{Hell96b}.  It comes in 3 varieties: quasi-spherical, quasi-planar, and quasi-hyperboloidal.  Loosely speaking, the metric can be described as being composed of a sequence of constant-$r$ shells (spherical, planar, or right hyperboloidal) that are arranged non-symmetrically relative to each other, and each of which evolves according to its own set of parameters, as determined by the local values of the arbitrary functions.  The arbitrary functions $S(r)$, $P(r)$ and $Q(r)$ determine how non-symmetric the arrangement of shells is, and the arbitrary functions $f(r)$, $M(r)$ and $a(r)$ determine the time evolution of each shell.  To be specific, the line element is
 \begin{align}
   ds^2 = - \d t^2 + \frac{\left( R' - R \frac{E'}{E} \right)^2 \, \d r^2}{\epsilon + f}
   + \frac{R^2}{E^2} \left( dp^2 + dq^2 \right) ~.   \showlabel{Szds2}
 \end{align}
 where
 $f = f(r)$ is a curvature-energy function, $R = R(t,r)$ obeys
 \begin{align}
   \Rt^2 = \frac{2 M}{R} + f + \frac{\Lambda R^2}{3} ~, \showlabel{RtSq}
 \end{align}
 and $M = M(r)$ is a mass-like function appearing in the ``gravitational potential" term of \er{RtSq}.  For quasi-spherical regions, it is the gravitational mass within the spherical shell of constant $r$.  The integration of \er{RtSq} introduces the bang time function $a(r)$ which defines the time on each constant $r$ shell when $R(a(r),r) = 0$.  The 3 arbitrary functions $f$, $M$ and $a$ and the $R(t,r)$ that depends on them are mathematically the same as in the \LT\ (LT) metric \cite{Lem33,Tol34}.  
 
   The evolution of $R$ depends on the value of $f$, or more correctly, the value of $f/M^{2/3}$.  When $\Lambda = 0$ the solutions of (\ref{RtSq}), in terms of a parameter $\eta$, are \\[1mm]
 {\bf Hyperbolic} ($f > 0$, $0 < \eta < \infty$)
 \begin{align}
   R & = \frac{M}{f}(\cosh\eta - 1) ~,   \showlabel{hyp0}\\
   (\sinh\eta - \eta) &= \frac{f^{3/2}(t - a)}{M} ~,   \showlabel{hyp1}
 \end{align}
 {\bf Parabolic} ($f = 0$)
 \begin{align}
   R & = \left(\frac{9M(t - a)^2}{2} \right)^{1/3} ~,   \showlabel{par1}
 \end{align}
 {\bf Elliptic} ($f < 0$, $0 < \eta < 2 \pi$)
 \begin{align}
   R & = \frac{M}{(-f)}(1 - \cos\eta) ~,   \showlabel{elip0}\\
   (\eta - \sin\eta) & = \frac{(-f)^{3/2}(t - a)}{M} ~,   \showlabel{elip1}
 \end{align} 
 and the time-reverse solutions are also possible.
 
   The function $E = E(r,p,q)$ is given by
 \begin{align}
   E = \frac{S}{2} \left( \frac{(p - P)^2}{S^2} + \frac{(q - Q)^2}{S^2} + \epsilon \right)   \showlabel{SzEdef}
 \end{align}
 and $\epsilon$ determines the geometry of the $(p-q)$ 2-surfaces; $+1$ for spherical, $0$ for planar, and $-1$ for hyperboloidal.  In fact a single metric can contain shells of all 3 types.  The coordinates $(p, q)$ are related to more usual coordinates on a sphere or hyperboloid, $(\theta, \phi)$, by a Riemann projection, see e.g. \cite{Hell09}.  However, in the latter coodinates the metric is not diagonal%
 \footnote{\sf Note that the observer's sky angles $\vth$ \& $\vph$ are different from the Szekeres coordinate angles $\theta$, $\phi$ defined via a Riemann projection, see \cite{Hell09}.}%
 .

   Although the terms ``radius" and ``radial" properly apply only in spherical symmetry, we will continue to describe $r$ as a ``radial" coordinate with all $\epsilon$ values, for lack of better terminology%
 \footnote{\sf There will be no detailed geometric discussion in this paper, within which a disctinction would be necessary.}%
 .  A key difference though is that quasi-planar and quasi-hyperboloidal models cannot have origins \cite{HelKra08}.

   The matter content is comoving dust, $T^{ab} = \rho U^a U^b$, $U^c = \delta^c_t$, for which the density is
 \begin{align}
   \kappa \rho = \frac{2 \left( M' - 3 M \frac{E'}{E} \right)}{R^2 \left(R' - R \frac{E'}{E} \right)} ~.
 \end{align}
 The comoving requirement breaks down and singularities form, when shells of matter at different $r$ values intersect.  Thus, to ensure a regular S model, one needs to check the conditions for no shell crossings \cite{HelLak85,HelKra02,HelKra08} are satisfied by the chosen arbitrary functions; these are summarised in \cite{Hell09}.

   The S metric contains the Ellis metrics \cite{Elli67} as a special case, when $S$, $P$ and $Q$ are constant, the LT metric being the $\epsilon = +1$ Ellis metric, and the Ellis metrics contains the dust Friedmann-\L-Robertson-Walker (FLRW) metric as the homogeneous special case.  Interestingly, the Szekeres case when $f(r)$, $M(r)$ and $a(r)$ take their FLRW forms, 
 \begin{align}
   M \propto f^{3/2} ~,~~~~~~ a = 0 ~,   \showlabel{scRWg}
 \end{align}
 while $S(r)$, $P(r)$ and $Q(r)$ are free (within the no shell crossing conditions), is also the FLRW metric in non-symmetric coordinates%
 \footnote{\sf The first statement of this seems to be by Goode and Wainwright in \cite{GooWai82}, where they remark at eq (2.24) that zero shear, and hence FLRW, is obtained with $\beta_+ = 0 = \beta_-$.  In the present notation this is $-{\rm sign}(f) \sqrt{|f|}\;/3\{M'/M - 3f'/(2f)\} = 0 = |f|^2 a'/(6 M)$, which does imply $M \propto f^{3/2}$ \& $a' = 0$.  Their notation does not allow $f$ to be zero away from an origin, $M = 0$.  See also section 1.3.4 of \cite{Kras97}.}%
 .  The Schwarzschild metric is obtained when $M$, $S$, $P$ \& $Q$ are constant, though care is needed to get good coordinate coverage \cite{HelKra02,Hell87}.

   One may think of a Szekeres model as an LT or Ellis model with additional inhomogeneities determined by the dipole functions $S$, $P$ \& $Q$.  This is because the evolution of $R(t,r)$, upon which the evolution of everything else depends, is completely determined by the three functions $f$, $M$ and $a$.  Therefore it is convenient to group the arbitrary functions into the ``LT functions" $f$, $M$ \& $a$, and the ``dipole functions" $S$, $P$ \& $Q$.

   See \cite{HelKra02} and \cite{HelKra08} for a discussion of the properties of the S metrics, \cite{WalHel12} or \cite{Hell17} for a list of citations, and \cite{Hell09} for a well-illustrated summary.  See also \cite{Kras97} for an extensive review of this and other inhomogeneous cosmologies.

 \subsection{The Observer Basis DEs}

   We place the observer at an arbitrary $(t,r,p,q)$ position, and for this implementation we assume both observer and emitter are comoving.  At the observer $O$, an orthonormal basis for \er{Szds2}, at a general location, is 
 \begin{align}
   \eo_0{}^a = \begin{pmatrix}  1 \\ 0 \\ 0 \\ 0  \end{pmatrix} ~,~~~~~~
   \eo_1{}^a & = \begin{pmatrix}  0 \\ \frac{\sqrt{\epsilon + f}\;}{\left( R' - \frac{R E'}{E} \right)}
      \\ 0 \\ 0  \end{pmatrix} ~,~~~~~~
   \eo_2{}^a = \begin{pmatrix}  0 \\ 0 \\ \frac{E}{R} \\ 0  \end{pmatrix} ~,~~~~~~
   \eo_3{}^a = \begin{pmatrix}  0 \\ 0 \\ 0 \\ \frac{E}{R}  \end{pmatrix} ~,
   \showlabel{SzONT}
 \end{align}
 and the observer's angles are shown relative to these in fig \ref{SzONTpic}.
 \\
 \psset{unit=1.4mm, xunit=1.4mm, yunit=1.4mm}
 \pspicture*(-35,-20)(90,23)     
 \psset{linewidth=0.7pt,linecolor=black,linearc=0.1}
 \psline[linestyle=dashed](9.54506,11.3754)(12.6821,5.00059)(2.74885,-6.78268)(0,0)
 \psline{->}(0,0)(16.7101,6.58882)
 \rput[lb](17.2,7.1){$\bk$}
 \psline(-7.60319,3.72014)(-7.45441,3.56043)(-7.30292,3.39936)(-7.14878,3.23701)(-6.99203,3.07343)
(-6.83273,2.90868)(-6.67094,2.74281)(-6.50671,2.5759)(-6.34012,2.408)(-6.17121,2.23917)
(-6.00004,2.06947)(-5.82669,1.89897)(-5.6512,1.72773)(-5.47365,1.55581)(-5.2941,1.38327)
(-5.11262,1.21019)(-4.92927,1.03661)(-4.74411,0.86262)(-4.55723,0.688267)(-4.36868,0.513621)
(-4.17854,0.338747)(-3.98687,0.163709)(-3.79375,-0.0114273)(-3.59924,-0.186596)(-3.40343,-0.361733)
(-3.20638,-0.536773)(-3.00816,-0.711649)(-2.80885,-0.886298)(-2.60853,-1.06065)(-2.40727,-1.23465)
(-2.20514,-1.40822)(-2.00221,-1.58131)(-1.79857,-1.75384)(-1.5943,-1.92576)(-1.38945,-2.09699)
(-1.18413,-2.26748)(-0.978385,-2.43716)(-0.772312,-2.60597)(-0.565982,-2.77384)(-0.359471,-2.94072)
(-0.152859,-3.10653)(0.0537801,-3.27123)(0.260367,-3.43474)(0.466826,-3.59701)(0.67308,-3.75798)
(0.879051,-3.91758)(1.08466,-4.07576)(1.28984,-4.23247)(1.49451,-4.38763)(1.69859,-4.5412)(1.902,-4.69312)
 \psline(8.90561,10.6133)(8.94296,10.5314)(8.97724,10.4459)(9.00845,10.3566)(9.03658,10.2636)
(9.0616,10.167)(9.0835,10.0668)(9.10227,9.96301)(9.11791,9.85566)(9.13039,9.74478)
(9.13971,9.63041)(9.14587,9.51259)(9.14885,9.39136)(9.14866,9.26675)(9.14528,9.1388)
(9.13872,9.00755)(9.12896,8.87305)(9.11602,8.73534)(9.09989,8.59446)(9.08057,8.45047)
(9.05807,8.3034)(9.03239,8.15331)(9.00353,8.00025)(8.97151,7.84426)(8.93632,7.6854)
(8.89798,7.52373)(8.8565,7.3593)(8.81188,7.19217)(8.76415,7.02238)(8.71332,6.85001)
(8.65939,6.67511)(8.60239,6.49774)(8.54233,6.31796)(8.47924,6.13584)(8.41312,5.95143)
(8.34401,5.76482)(8.27192,5.57605)(8.19688,5.3852)(8.11891,5.19233)(8.03805,4.99751)
(7.9543,4.80082)(7.86771,4.60232)(7.7783,4.40209)(7.6861,4.20019)(7.59115,3.9967)
(7.49348,3.79169)(7.39311,3.58524)(7.29009,3.37742)(7.18445,3.1683)(7.07623,2.95797)(6.96546,2.7465)
 \rput(6.9,6){$\vth$}
 \rput(-3.7,-2.8){$\vph$}
 \psset{linewidth=1.2pt,linecolor=blue}
 \psline{->}(0,0)(15.9642,19.0253)
 \rput[lb](16.5,19.5){$\beo_3 = \beo_q$}
 \psline{->}(0,0)(14.4966,-15.509)
 \rput[lt](15,-16){$\beo_2 = \beo_p$}
 \psline{->}(0,0)(-13.8282,6.76597)
 \rput[rb](-14.3,7.3){$\beo_1 = \beo_r$}
 \rput[lt](28,5){
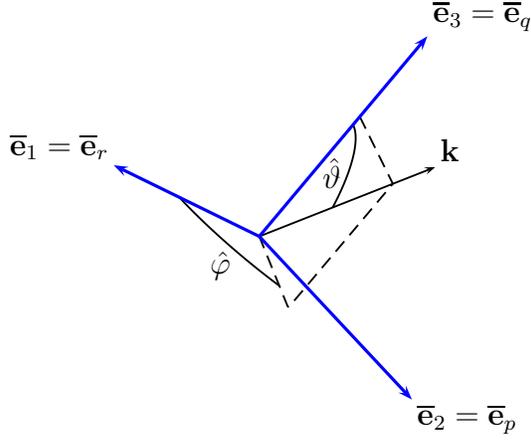
   \pb{75mm}{\sf \small \Fgr{SzONTpic} The observer's ray angles and reference directions defined for the Szekeres metric.}
 }
 \endpspicture

   For any given initial ray direction, $(\vth, \vph)$, we find the initial PNC basis vectors by inserting \er{OthNmBbar} into \er{NlRdBhat}, using \er{SzONT} for the components of $\eo_i{}^a$, and taking the $\chh \to 0$ limit.  The result agrees with \er{etahIC}, \er{evthIC} \& \er{evphIC}:
 \begin{align}
   \eh_\tah{}^a & = \begin{pmatrix}  1 \\ 0 \\ 0 \\ 0  \end{pmatrix} ~,~~ &
   \eh_\chh{}^a & = \begin{pmatrix}
      -1 \\
      \sin\vth \cos\vph \frac{\sqrt{\epsilon + f}\;}{\left( R' - \frac{R E'}{E} \right)} \\
      \sin\vth \sin\vph \frac{E}{R} \\
      \cos\vth \frac{E}{R}
             \end{pmatrix} ~, \nn \\
   \eh_\vth{}^a & = \begin{pmatrix}
      0 \\
      \chh \cos\vth \cos\vph \frac{\sqrt{\epsilon + f}\;}{\left( R' - \frac{R E'}{E} \right)} \\
      \chh \cos\vth \sin\vph \frac{E}{R} \\
      - \chh \sin\vth \frac{E}{R}
             \end{pmatrix} ~~~~~~~~\to &
   \eh_\vth{}^a \Big|_{\chh = 0} & = \begin{pmatrix} 0 \\ 0 \\ 0 \\ 0 \end{pmatrix} ~, \nn \\
   \eh_\vph{}^a & = \begin{pmatrix}
      0 \\
      - \chh \sin\vth \sin\vph \frac{\sqrt{\epsilon + f}\;}{\left( R' - \frac{R E'}{E} \right)} \\
      \chh \sin\vth \cos\vph \frac{E}{R} \\
      0
             \end{pmatrix} ~~~~~~~~\to &
   \eh_\vth{}^a \Big|_{\chh = 0} & = \begin{pmatrix} 0 \\ 0 \\ 0 \\ 0 \end{pmatrix} ~.
   \showlabel{SzNRBh}
 \end{align}
 Note that, by \er{inik} the initial $k^a$ is the $\eh_\chh{}^a$ given in \er{SzNRBh}.  Their initial derivatives are found by using \er{SzONT} in \er{detahIC}, \er{devthIC} \& \er{devphIC},
 \begin{align}
   \td{}{\chh} \eh_\tah{}^a \Bigg|_{\chh = 0} & =
      \begin{pmatrix}
      0 \\
      \dfrac{- \sqrt{\epsilon + f}\; (\Rt' - \Rt E'/E)}{(R' - R E'/E)^2} \sin\vth \cos\vph \\
      \dfrac{- \Rt E}{R^2} \sin\vth \sin\vph \\
      \dfrac{- \Rt E}{R^2} \cos\vth
      \end{pmatrix}_o ~, \nn \\
   \td{}{\chh} \eh_\chh{}^a & =
      \begin{pmatrix}
      - \sin^2\vth \cos^2\vph \left( \frac{(\Rt' - \Rt E'/E)}{(R' - R E'/E)} - \frac{\Rt}{R} \right) - \frac{\Rt}{R} \\
      \end{pmatrix}_o ~, \nn \\
   \td{}{\chh} \eh_\vth{}^a \Bigg|_{\chh = 0} & =
      \begin{pmatrix}
      0 \\
      \cos\vth \cos\vph \frac{\sqrt{\epsilon + f}\;}{(R' - R E'/E)} \\
      \cos\vth \sin\vph \frac{E}{R} \\
      - \sin\vth \frac{E}{R}
      \end{pmatrix}_o ~, \nn \\
   \td{}{\chh} \eh_\vph{}^a \Bigg|_{\chh = 0} & =
      \begin{pmatrix}
      0 \\
      - \sin\vth \sin\vph \frac{\sqrt{\epsilon + f}\;}{(R' - R E'/E)} \\
      \sin\vth \cos\vph \frac{E}{R} \\
      0
      \end{pmatrix}_o
   \showlabel{Sz-dedchi}
 \end{align}
 and by using \er{SzONT} in \er{kNullGeod}
 \begin{align}
   \td{}{\chh} \eh_\chh{}^t \Big|_{\chh = 0} & = \Bigg[
         - \left( \frac{(\Rt' - \Rt E'/E)}{(R' - R E'/E)} - \frac{\Rt}{R} \right) \sin^2\vth \cos^2\vph - \frac{\Rt}{R}
         \Bigg]_o \\
   \td{}{\chh} \eh_\chh{}^r \Big|_{\chh = 0} & = \Bigg[ \frac{\sqrt{\epsilon + f}\;}{(R' - R E'/E)} \Bigg\{
         \frac{2 (\Rt' - \Rt E'/E)}{(R' - R E'/E)} \sin\vth \cos\vph \nn \\
      &~~~~ + \frac{2 E}{(R' - R E'/E)} \sin\vth \cos\vph
         \left[ \left( \frac{E'_p}{E} - \frac{E' E_p}{E^2} \right) \sin\vth \sin\vph
         + \left( \frac{E'_q}{E} - \frac{E' E_q}{E^2} \right) \cos\vth \right] \Bigg\} \nn \\
      &~~~~ + \frac{(\epsilon + f)}{(R' - R E'/E)} \Bigg\{
         \frac{1}{R} (1 - \sin^2\vth \cos^2\vph) \nn \\
      &~~~~ - \frac{(R'' - (R' E' + R E'')/E + R E'^2/E^2)}{(R' - R E'/E)^2} \sin^2\vth \cos^2\vph \Bigg\} \nn \\
      &~~~~ + \frac{2 f'}{(R' - R E'/E)^2} \sin^2\vth \cos^2\vph \Bigg]_o \\
   \td{}{\chh} \eh_\chh{}^p \Big|_{\chh = 0} & = \Bigg[
         - \frac{2 E \sqrt{\epsilon + f}\;}{R^2} \sin^2\vth \cos\vph \sin\vph
         + \frac{2 \Rt E}{R^2} \sin\vth \sin\vph
         + \frac{2 E E_q}{R^2} \sin\vth \sin\vph \cos\vth \nn \\
      &~~~~ + \frac{E E_p}{R^2} (\sin^2\vth \sin^2\vph  - \cos^2\vth)
         - \frac{E^2}{R B} \left( \frac{E'_p}{E} - \frac{E' E_p}{E^2} \right) \sin^2\vth \cos^2\vph \Bigg]_o \\
   \td{}{\chh} \eh_\chh{}^q \Big|_{\chh = 0} & = \Bigg[
         - \frac{2 E \sqrt{\epsilon + f}\;}{R^2} \sin\vth \cos\vth \cos\vph
         + \frac{2 \Rt E}{R^2} \cos\vth
         + \frac{2 E E_p}{R^2} \sin\vth \cos\vth \sin\vph \nn \\
      &~~~~ - \frac{E E_q}{R^2} (\sin^2\vth \sin^2\vph  - \cos^2\vth)
         - \frac{E^2}{R B} \left( \frac{E'_q}{E} - \frac{E' E_q}{E^2} \right) \sin^2\vth \cos^2\vph \Bigg]_o
 \end{align}

   The commutation coefficients, the Christoffel symbols and their partial derivatives are calculated using GRTensor \cite{GRT} \& Maple \cite{Maple}.  See appendix \ref{Cmt} for the commutation coefficients.

 \subsection{Area Distance and Redshift Space Number Density}

   When calculating the area elements and area distance of eqs \er{dAreaWedge}, \er{dArea} \& \er{dA2}, for the Szekeres model we choose comoving emitters,
 \begin{align}
   \th^c & = U^c = \delta^c_t ~,~~~~~~ \sqrt{|g|}\; = \frac{\left( R' - \frac{R E'}{E} \right) R^2}{\sqrt{\epsilon + f}\; E^2}
      ~,~~~~~~ C = k^t ~~~~~~~~\to  \\
   d_A^2 & = \left| \frac{\sqrt{|g|}\;}{k^t \sin\vth} \, \varepsilon_{tbcd} \, k^b \, \eh_\vth{}^c \, \eh_\vph{}^d
      \right| \\
   & = \left| \frac{\sqrt{|g|}\;}{k^t \sin\vth} \Big(
      k^r \, \eh_\vth{}^p \, \eh_\vph{}^q - k^r \, \eh_\vth{}^q \, \eh_\vph{}^p
      + k^p \, \eh_\vth{}^q \, \eh_\vph{}^r - k^p \, \eh_\vth{}^r \, \eh_\vph{}^q
      + k^q \, \eh_\vth{}^r \, \eh_\vph{}^p - k^q \, \eh_\vth{}^p \, \eh_\vph{}^r
      \Big) \right| ~.
      \showlabel{SzdA}
 \end{align}

   Similarly for the redshift space number density we get
 \begin{align}
   \nh & = \frac{2 \left( M' - 3 M \frac{E'}{E} \right)}{\kappa R^2 \left(R' - R \frac{E'}{E} \right)}
      \Bigg| \frac{\sqrt{|g|}\;}{\mu \, \left( \td{}{\chh} \eh^\tah{}_t \right) \, \sin\vth} \times \nn \\
   &~~~~ \Big(
      k^r \, \eh_\vth{}^p \, \eh_\vph{}^q - k^r \, \eh_\vth{}^q \, \eh_\vph{}^p
      + k^p \, \eh_\vth{}^q \, \eh_\vph{}^r - k^p \, \eh_\vth{}^r \, \eh_\vph{}^q
      + k^q \, \eh_\vth{}^r \, \eh_\vph{}^p - k^q \, \eh_\vth{}^p \, \eh_\vph{}^r
      \Big)      \Bigg| ~.
      \showlabel{Sznh}
 \end{align}
 Eq \er{nhat} allows numerical computation of the observable number density in redshift space, given the metric quantities and $\mu$.

   This section has presented the specialisation of our general framework to the Szekeres inhomogeneous cosmology with comoving emitters and observers.  In Paper II, \cite{WalHel17}, we shall develop the general framework and Szekeres example into a numerical algorithm coded in MATLAB.

 \section{Conclusions}

   Cosmological observations are becoming ever more precise, and the interpretation of those observations needs increasingly detailed modelling.  Amongst the available tools, only `inhomogeneous cosmologies' are both exact and fully non-linear.  Thus understanding the relation between exact inhomogeneities and observations is fast becoming indispensible.

   We have set up a clear and simple framework for calculating a set of cosmological observables, for a randomly placed observer, watching arbitrarily moving sources, in a general inhomogeneous spacetime, assuming the metric is given.  The framework is based on the null geodesic equation and the geodesic deviation equation.  The standard tensor form of these equations, ideal for mathematical manipulation, was recast in terms of total derivatives to make the equations suitable for a numerical treatment.  The framework consists of setting up a past null tetrad for an observer with arbitrary position and motion, and then Lie dragging the observer's angle and time coordinates down her past null cone.  This results in a numerically calculated coordinate basis, and this basis allows the easy calculation of the redshift and proper motion of any source along the line of sight, as well as the area distance and redshift space number density.  Other variables will be considered in future.  The discussion in \cite{MuHeEl97} and in \cite{Hel01} regarding source evolution and the use of multicolour observations apply here too.

   Our framework is applicable to general observers and emitters in any given spacetime metric, and it combines generality with relative simplicity and numerical efficiency, so it is a valuable tool in its own right.  Furthermore, this approach is complementary to the `Metric of the Cosmos' project, that seeks to derive the metric from observations, in that our framework  can provide test data.

   A problem often encountered in tracing light rays between source and observer in non-symmetric spacetimes is that rays from the same emitter at successive times are emitted in different directions and arrive at the observer from different directions.  Our formalism provides a simple solution to this problem, since the instantaneous rate of change of observed angle is obtained directly from the proper motion.

   The essential results from this paper, that a numerical realisation will need, are the PNC definition \er{NlRdBhat}, the numerical null geodesic equation and initial condition \er{kNullGeod} \& \er{inik}, the numerical basis propagation (geodesic deviation) equation \er{GeodDevTot}/\er{WvvUp} and its alternative for propagating inverse basis \er{WvvDn}, the initial conditions for basis propagation \er{etahIC}, \er{evthIC}, \er{evphIC}, \er{detahIC}, \er{devthIC} and \er{devphIC}, the redshift \er{redshift}, the components of proper motion \er{appmotth}, \er{appmotph}, the area distance \er{dA2} with \er{dArea} or \er{dAreaWedge}, and the redshift space density \er{nhat}.  The basis tilt equations \er{edotvt-sin} could also be useful.  In addition, for any given metric, such as the Szekeres example above, one would need a choice of ONT \er{SzONT}, and the specific application of the above \er{SzNRBh}-\er{Sznh}.

   Paper II will describe the numerical implementation of this framework, and begin the process of exploring observations in a variety of cosmological models of different scales.


 \appendix

 \section{Propagating the Observer's Basis Using the Geodesic Deviation Equation}
 \showlabel{GDprop}

 \subsection{Derivation}

   The standard geodesic deviation equation uses tensor derivatives, which are ideal for physical understanding and for doing covariant calculations.  But to actually integrate vector or tensor components along a path, we need to convert the absolute derivatives into total derivatives, and re-write the equation as an ordinary differential equation.

   ${}$ \indent The geodesic deviation equation is 
 \begin{align}
   \adt{W^a}{v} & = - R^a{}_{bcd} V^b W^c V^d ~,
      \showlabel{GD} \\
   R^a{}_b{}_{cd} & = - \Gamma^a{}_{cb,d} + \Gamma^a{}_{db,c}
      - \Gamma^e{}_{cb} \, \Gamma^a{}_{de} + \Gamma^e{}_{db} \, \Gamma^a{}_{ce} ~,
      \showlabel{RiemGam}
 \end{align}
 where $W^a$ is the deviation vector field, $V^a = \tdil{x^a}{v}$ is the vector field for a geodesic congruence, and $v$ is the affine parameter along $V^a$, while \er{RiemGam} applies to a coordinate basis.  The requirements for the geodesic deviation construction are that $V^c$ is geodesic, and it commutes with $W^d$,
 \begin{align}
   \ad{V^a}{v} = \td{V^a}{v} +  V^b \Gamma^a{}_{bc} V^c & = 0 = V^b \nabla_b V^a
      \showlabel{Vg} \\
   V^b \nabla_b W^a - W^b \nabla_b V^a & = 0
 \end{align}
 However, in using \er{GD} to propagate $W^c$ along a geodesic congruence, we actually need the variation of the components of $W^b$, that is $\tdtil{W^a}{v}$.  Thus 
 \begin{align}
   \ad{W^a}{v} & = V^b \big( \p_b W^a + \Gamma^a{}_{bc} W^c \big) \\
   & = \td{W^a}{v} + V^b \Gamma^a{}_{bc} W^c \\
   \ad{}{v} \ad{W^a}{v} & = V^b \left( \p_b \ad{W^a}{v} + \Gamma^a{}_{bc} \ad{W^c}{v} \right) \\
   & = V^b \left( \p_b \left\{ \td{W^a}{v} + V^d \Gamma^a{}_{dc} W^c \right\}
      + \Gamma^a{}_{bc} \left\{ \td{W^c}{v} + V^d \Gamma^c{}_{de} W^e \right\} \right) \\
   & = \left\{ \tdt{W^a}{v} + \td{V^d}{v} \Gamma^a{}_{dc} W^c + V^b V^d \Gamma^a{}_{dc,b} W^c
      + V^d \Gamma^a{}_{dc} \td{W^c}{v} \right\} \nn \\
   &~~~~~~ + \left\{ V^b \Gamma^a{}_{bc} \td{W^c}{v} + V^b \Gamma^a{}_{bc} V^d \Gamma^c{}_{de} W^e \right\} \\
   & = \tdt{W^a}{v} + \Gamma^a{}_{bc} \left( W^c \td{V^b}{v} + 2 V^b \td{W^c}{v} \right)
      + V^b W^c V^d \big( \Gamma^a{}_{dc,b} + \Gamma^a{}_{be} \, \Gamma^e{}_{dc} \big) \\
   \mbox{(by \er{GD})}~~~~~~ & = - R^a{}_{bcd} V^b W^c V^d
      = V^b W^c V^d \big( \Gamma^a{}_{cb,d} - \Gamma^a{}_{db,c}
      + \Gamma^e{}_{cb} \, \Gamma^a{}_{de} - \Gamma^e{}_{db} \, \Gamma^a{}_{ce} \big) \\
   \therefore~~~~~~ \tdt{W^a}{v} & = - \Gamma^a{}_{bc} \left( W^c \td{V^b}{v}
      + 2 V^b \td{W^c}{v} \right) \nn \\
   &~~~~~~ + V^b W^c V^d \big( \Gamma^a{}_{cb,d} - \Gamma^a{}_{db,c} - \Gamma^a{}_{dc,b}
      + \Gamma^e{}_{cb} \, \Gamma^a{}_{de} - \Gamma^e{}_{db} \, \Gamma^a{}_{ce}
      - \Gamma^a{}_{be} \, \Gamma^e{}_{dc} \big) \\
   & = - \Gamma^a{}_{bc} \left( W^c \td{V^b}{v} + 2 V^b \td{W^c}{v} \right) 
      - V^b W^c V^d \big( \Gamma^a{}_{db,c} + \Gamma^e{}_{db} \, \Gamma^a{}_{ce} \big) \\
   & = - 2 \Gamma^a{}_{bc} V^b \td{W^c}{v} - V^b W^c V^d \Gamma^a{}_{db,c}
      - \Gamma^a{}_{ec} W^c \left( \td{V^e}{v} + V^b V^d \Gamma^e{}_{db} \right) \\
   \mbox{(by \er{Vg})}~~~~~~ \tdt{W^a}{v}& = - V^b \left( 2 \Gamma^a{}_{bc} \td{W^c}{v}
      + W^c V^d \Gamma^a{}_{db,c} \right)
      \showlabel{WvvUp}
 \end{align}

   For completeness, we repeat the calculation for a covariant $W_b$, since $g_{ab}$ does not commute with $\tdil{}{v}$:
 \begin{align}
   \ad{W_a}{v} & = V^b (\p_b W_a - \Gamma^c{}_{ba} W_c) \\
   & = \td{W_a}{v} - V^b \Gamma^c{}_{ba} W_c \\
   \ad{}{v} \ad{W_a}{v} & = V^b \left( \p_b \ad{W_a}{v} - \Gamma^c{}_{ba} \ad{W_c}{v} \right) \\
   & = V^b \left( \p_b \left\{ \td{W_a}{v} - V^d \Gamma^c{}_{da} W_c \right\}
      - \Gamma^c{}_{ba} \left\{ \td{W_c}{v} - V^d \Gamma^e{}_{dc} W_e \right\} \right) \\
   & = \left\{ \tdt{W_a}{v} - \td{V^d}{v} \Gamma^c{}_{da} W_c - V^b V^d \Gamma^c{}_{da,b} W_c
      - V^d \Gamma^c{}_{da} \td{W_c}{v} \right\} \nn \\
   &~~~~~~ - \left\{ V^b \Gamma^c{}_{ba} \td{W_c}{v} - V^b \Gamma^c{}_{ba} V^d \Gamma^e{}_{dc} W_e \right\} \\
   & = \tdt{W_a}{v} - \Gamma^c{}_{ba} \left( W_c \td{V^b}{v} + 2 V^b \td{W_c}{v} \right)
      - V^b W_c V^d \big( \Gamma^c{}_{da,b} - \Gamma^e{}_{ba} \, \Gamma^c{}_{de} \big) \\
   & = - R_{ab}{}^c{}_d V^b W_c V^d = - R^c{}_{dab}{} V^b W_c V^d \\
   & = V^b W_c V^d \big( \Gamma^c{}_{ad,b} - \Gamma^c{}_{bd,a}
      + \Gamma^e{}_{ad} \, \Gamma^c{}_{be} - \Gamma^e{}_{bd} \, \Gamma^c{}_{ae} \big) \\
   \tdt{W_a}{v} & = + \Gamma^c{}_{ba} \left( W_c \td{V^b}{v} + 2 V^b \td{W_c}{v} \right)
      + V^b W_c V^d \big( \Gamma^c{}_{da,b} + \Gamma^c{}_{ad,b} - \Gamma^c{}_{bd,a} \nn \\
   &~~~~~~ - \Gamma^e{}_{ba} \, \Gamma^c{}_{de} + \Gamma^e{}_{ad} \, \Gamma^c{}_{be}
      - \Gamma^e{}_{bd} \, \Gamma^c{}_{ae} \big) \\
   \mbox{(by \er{Vg})}~~~~~~ & = \Gamma^c{}_{ba} \left( W_c \left\{ - V^e V^d \Gamma^b{}_{ed} \right\}
      + 2 V^b \td{W_c}{v} \right) \nn \\
   &~~~~~~ + V^b W_c V^d \big( 2 \Gamma^c{}_{ad,b} - \Gamma^c{}_{bd,a}
      - \Gamma^e{}_{bd} \, \Gamma^c{}_{ae} \big) \\
   \tdt{W_a}{v} & = V^b \left( 2 \Gamma^c{}_{ba} \td{W_c}{v}
      + W_c V^d \big( 2 \Gamma^c{}_{ad,b} - \Gamma^c{}_{bd,a}
      - 2 \Gamma^c{}_{ea} \, \Gamma^e{}_{bd} \big) \right)
      \showlabel{WvvDn}
 \end{align}

 \subsection{Consistency of \er{kNullGeod} \& \er{GeodDevTot}}
 \showlabel{ConsistencyGeodDev}

   We here check the basis propagation equation \er{WvvUp} with an obvious special case: the propagation of $\eh_\chi^a = k^a$ should be consistent with the geodesic equation.  We differentiate the radial geodesic equation \er{kNullGeod} to get
 \begin{align}
   \td{k^a}{\chi} & = - \Gamma^a{}_{bc} k^b k^c \\
   \tdt{k^a}{\chi} & = - \td{}{\chi} \Big( \Gamma^a{}_{bc} k^b k^c \Big)
      = - k^d \p_d \Big( \Gamma^a{}_{bc} k^b k^c \Big) \\
   & = - k^d \big( \Gamma^a{}_{bc,d} k^b k^c + 2 \Gamma^a{}_{bc} k^b \p_d k^c \big) \\
   & = - k^d \Gamma^a{}_{bc,d} k^b k^c - 2 k^d \Gamma^a{}_{bc} k^b \big( \nabla_d k^c - \Gamma^c{}_{de} k^e \big) \\
   & = - k^d \Gamma^a{}_{bc,d} k^b k^c + 2 k^d \Gamma^a{}_{bc} k^b \Gamma^c{}_{de} k^e \\
   & = - k^d \Gamma^a{}_{bc,d} k^b k^c + 2 k^d \Gamma^a{}_{be} k^b \Gamma^e{}_{dc} k^c \\
   & = - k^b k^c k^d \Big( \Gamma^a{}_{bc,d} - 2 \Gamma^a{}_{be} \Gamma^e{}_{dc} \Big) ~;   \showlabel{Ck1}
 \end{align}
 while \er{WvvUp}, with the substitutions $V^a = k^a = W^a$ \& $v = \chi$ --- that is \er{GeodDevTot}, becomes
 \begin{align}
   \mbox{(by \er{Vg})}~~~~~~ \tdt{k^a}{\chi} 
      & = - k^b \Big( 2 \Gamma^a{}_{bc} \td{k^c}{\chi} + k^c k^d \Gamma^a{}_{db,c} \Big) \\
   & = - 2 k^b \Gamma^a{}_{be} \td{k^e}{\chi} - k^b k^c k^d \Gamma^a{}_{db,c} \\
   & = - 2 k^b \Gamma^a{}_{be} \Big( - \Gamma^e{}_{dc} k^d k^c \Big) - k^b k^c k^d \Gamma^a{}_{db,c} \\
   & = + 2 k^b \Gamma^a{}_{be} \Gamma^e{}_{dc} k^d k^c - k^c k^d k^b \Gamma^a{}_{bc,d} \\
   & = k^b k^c k^d \Big( 2 \Gamma^a{}_{be} \Gamma^e{}_{dc} - \Gamma^a{}_{bc,d} \Big) ~.   \showlabel{Ck2}
 \end{align}
 The agreement of \er{Ck1} \& \er{Ck2} supports the validity of \er{WvvUp}.

 \section{Commutators}
 \showlabel{Cmt}

   If the propagated basis $\beh_\alpha$ is to be a coordinate basis, then its commutators must all be zero.  By construction, that is by \er{ekLie} and \er{dedchiLie}, we already have all the $[\beh_\chh, \beh_\alpha] = 0$ for $\beh_\chh = {\bf k}$ and all $\alpha$.  By the Jacobi identity, the remaining commutators are preserved by Lie dragging along the PNC --- that is, their values on $O$'s worldline are preserved.  Near $O$ we find
 \begin{align}
   [\beh_\tah, \beh_\vth] & = \chh \big( \cos\vth \cos\vph \, [\beo_\tao, \beo_1]
                            + \cos\vth \sin\vph \, [\beo_\tao, \beo_2]
                            - \sin\vth \, [\beo_\tao, \beo_3] \big)   \showlabel{heta-heth} \\
   [\beh_\tah, \beh_\vph] & = - \chh \big( \sin\vth \sin\vph \, [\beo_\tao, \beo_1]
                            + \sin\vth \cos\vph \, [\beo_\tao, \beo_2] \big)   \showlabel{heta-heph} \\
   [\beh_\vth, \beh_\vph] & = - \chh^2 \sin\vth \big( \cos\vth \, [\{\beo_\tao \! - \! \beo_1\}, \beo_2]
                            - \sin\vth \sin\vph \, [\beo_1, \beo_3]
                            + \sin\vth \cos\vph \, [\beo_2, \beo_3] \big)   \showlabel{heth-heph}
 \end{align}
 and
 \begin{align}
   [\beh_\tah, \beh_\chh] & = \sin\vth \cos\vph \, [\beo_\tao, \beo_1]
                            + \sin\vth \sin\vph \, [\beo_\tao, \beo_2]
                            + \cos\vth \, [\beo_\tao, \beo_3]   \showlabel{heta-hech} \\
   [\beh_\chh, \beh_\vth] & = \chh \big( \cos\vph \, [\beo_\tao, \beo_3]
                            - \sin\vph \, [\beo_2, \beo_3]
                            - \cos\vph \, [\beo_1, \beo_3] \big)   \showlabel{hech-heth} \\
   [\beh_\chh, \beh_\vph] & = \chh \sin\vth \big( - \sin\vth \, [\beo_\tao \! - \! \beo_1, \beo_3]
                            - \cos\vth \sin\vph \, [\beo_\tao \! - \! \beo_1, \beo_3]
                            - \cos\vth \cos\vph \, [\beo_2, \beo_3] \big)   \showlabel{hech-heph}
 \end{align}
 The first list need to be all zero (near $O$), but the second list do not, since they are not preserved by Lie dragging.  Note that \er{heta-heth}-\er{heth-heph} all go to zero on $O$ where $\chh = 0$, and \er{heth-heph} is second order in $\chh$.  The zero values of these commutators at $O$ are then preserved by the Lie dragging, so the constructed basis is a coordinate basis.
 Such a coordinate system can always be set up near a single worldline.  Indeed, there should be no problem setting up the orthonormal basis near $O$'s worldline so that all the commutators $[\be_a, \be_b]$ are locally zero to first order.

   For the Szekeres basis \er{SzONT} we find the non-zero commutators are
 \begin{align}
   &~~~~~~~~~~~~~~~~~~~~ [\beo_i, \beo_j] = \gmo^k{}_{ij} \beo_k \\
   & \left. \begin{matrix}
   \gmo^2{}_{02} = \dfrac{\Rt}{R} = \gmo^3{}_{03} ~,~~~~~~~~ &
      \gmo^1{}_{01} = \dfrac{\Rt' - \Rt E'/E}{R' - R E'/E} \\
   \gmo^2{}_{32} = \dfrac{E_q}{R} ~, &
      \gmo^1{}_{21} = \dfrac{E'_p - E_p E'/E}{R' - R E'/E} \\
   \gmo^3{}_{23} = \dfrac{E_p}{R} ~, &
      \gmo^1{}_{31} = \dfrac{E'_q - E_q E'/E}{R' - R E'/E}\\
   &
      \gmo^2{}_{21} = \dfrac{\sqrt{\epsilon + f}\,}{R} = \gmo^3{}_{31}
   \end{matrix}~~~~ \right\}
 \end{align}
 These are all finite at a general point, i.e. for a generic observer.  The only divergencies occur at a Szekeres ``origin", $R = 0$.  Should the observer pass through $R = 0$, we can take $\chh \sim R$ nearby, so the commutators of the observer's PNC coordinates remain zero.

 \end{document}